# *sp²* Amorphous carbons in view of multianalytical consideration: Normal, expected and new

Ye.A.Golubev[1], N.N.Rozhkova[2], E.N.Kabachkov[3,4], Yu.M.Shul'ga[3,5], K.Natkaniec-Holderna[6], I.Natkaniec[6], I.V.Antonets[7], B.A.Makeev[1], N.A.Popova[8], V.A.Popova[8], E.F.Sheka[8*]

[1] *Yushkin's Institute of Geology, Komi Science Center, Ural Branch of RAS, ul. Pervomayskaya, 54, 167982 Syktyvkar, Russia*

[2] *Institute of Geology, Karelian Research Centre RAS, ul. Pushkinskaya, 11, 185910 Petrozavodsk, Russia*

[3] *Institute of Problems of Chemical Physics RAS, pr. Academician Semenov, 1, 142432 Chernogolovka, Russia*

[4] *Scientific Centre RAS, ul. Lesnaya, 9, 142432 Chernogolovka, Russia*

[5] *Department of Dielectrics and Semiconductors, National University of Science and Technology MISIS, pr.Leninsky, 4, 119049 Moscow, Russia*

[6] *Faculty of Physics, Adam Mickiewicz University, ul. Umultowska, 85, 61-614 Poznań, Poland*

[7] *Department of Radiophysics, Syktyvkar State University, ul. Oktyabrskaya, 55, 167000 Syktyvkar, Russia*

[8] *Institute of Physical Research and Technology, Peoples' Friendship University of Russia, ul. Miklukho-Maklaya, 6, 117198 Moscow, Russia*

*corresponding author: sheka@icp.ac.ru

**Abstract:**

The suggested approach "Selected set of samples + selected set of analytical tools" occurred quite efficient when applying to *sp²* amorphous carbons, thus providing a transformation of 'amorphous' representation of the issue, based on particulars, into a 'crystalline' one based on a limited set of fixed commonalities. The slogan first part implies a set of different-origin solid samples. The second part concerns analytical tools, the most suitable to achieve the goal. The parts combining means the application of each tool to the whole set of samples. In the current study, two natural *sp²* amorphous carbons, namely, shungite carbon and antraxolite, as well as two engineered products – carbon blacks CB632 and CB624, all of the four bodies belonging to the elitist highest-carbon-content *sp²* species, were subjected to analytical study by using modern structural and compositional analytical techniques. The approach has allowed disclosing the following steady points that are common to the whole class of this carbon allotrope and that may lay the foundation of consolidate, more 'crystalline' representation of what are *sp²* amorphous carbons:

1. *sp²* Amorphous carbons are products of particular chemical reactions related to their basic structural units. Further macroscopic agglomeration of the latter plays a subsidiary role.
2. The units represent framed graphene molecules of 1-2 nm and 1- x*10 (x=1-3) nm in size in the case of natural and engineered products, respectively.

3. Framing of graphene molecules, predominantly incomplete with respect to the number of vacant places, concerns only edge atoms and is implemented by the related chemical additives, such as hydrogen, oxygen, nitrogen, sulfur and halogens which are attached to the carbon core via chemical bonding.
4. The molecules small size provides a countable number of newly formed chemical bonds. INS and XPS allow attributing the bonds to chemical compositions restricted by number while QCh ensures reliable support.
5. Temperature and pressure as well as physical state and chemical content of surrounding media are the main factors governing geochemistry and technical chemistry of carbon products.
6. Graphene molecules, laying the foundation of $sp^2$ amorphous carbons, are strongly radicalized due to which the latter acquire a new facet in the space of their properties, being the largest repository of stable radicals.

1. Introduction

Amorphous carbons are the most widely studied and practically used materials with the longest history of their investigation and exploitation. Their scientific understanding was developing alongside with growing material science, receiving at each turn of the progress a stimulating surge of a deeper penetration into understanding of their nature. Consequently, once started as classical continuous bulks, today's amorphous carbons unhesitatingly take their place among nanotechnological materials. However, despite the undoubtedly handy successes, a clear idea of what you are, amorphous carbons, is lost in an amorphous medium of almost countless number of individual facts and partial relations. And then the idea arose to use the power of modern tools used in nanotechnology in order to build a vision of amorphous carbon from the bottom. To a large extent this upward movement becomes possible due to the revolutionary development of analytical technologies, which led to such a dramatic change in modern materials knowledge as science.

Nanoscience and nanotechnology are still under constant growth and the importance of reliable characterization is evidently increased [1]. The multidisciplinary aspects of the fields do not permit every research team to have easy access to a broad range of characterization facilities. The acquisition of a full picture of the variety of features that are associated with a typical nanotechnological object requires the use of numerous techniques, often needing to use more than one of them for evaluating completely enough even a single property. At the same time it should be taken into account that the use of each modern technique is not simply analytical test but a quite sophisticated scientific problem due to which any presentation of the technique in a comparative way may act as a robust guide, helping to understand better its ability. Accordingly, the application of collected analytical tools for a selected group of samples has allowed not only to support the available knowledge as well as expected characteristics of the latter, but to reveal quite new faces of the matter significantly enlarging our presentation.

A large assortment of analytical capabilities on the service of nanotecnology undoubtedly greatly facilitates the choice in general, while every specific case should be considered separately depending on which namely characteristics of the studied matter are under question. In the current paper we consider the characterization of a selected set of representatives of a large class of $sp^2$ amorphic carbons (ACs) from the viewpoint of their detailed structure as well as chemical composition involving its description in terms of both wt% chemical content and assignment to specific functional groups forming the chemical structure. The following techniques aiming at determining structure and chemical composition has been selected.
1. High-resolved transmission electron microscopy (HRTEM) and scanning transmission electron microscopy (STEM)
2. X-ray powder diffraction (XRPD);
3. Thermal neutron powder diffraction (NPD);
4. Thermal gravimetric analysis (TGA);
5. Differential scanning calorimetry (DSC);
6. Combustion-based elemental analysis (EA);
7. Scanning electron microscopy (SEM) complemented with spectroscopic mapping by energy dispersive X-rays (EDX) (SEM-EDX) or energy dispersive spectroscopy (EDS) (SEM-EDS);
8. X-ray photoelectron spectroscopy (XPS);
9. Inelastic neutron scattering (INS);
10. Quantum-chemical calculations (QCh).

The first three techniques STEM/HRTEM, XRD and NPD form the ground of the species structure determination. TGA and DSC deal with the samples *brutto* (proximate) chemical content; therewith TGA concerns the sample mass and points to external inclusions (water and impurities) of the samples while DSC concerns chemical and structural characteristics which have the greatest effect on thermophysical properties of the samples. The elemental CHNS determination, provided by standard EA, is aimed at the determination of *netto* (ultimate) chemical content. SEM-EDS and XPS studies were mainly concentrated on the O/C composition of the samples. Previously performed INS played the role of internal control of the data obtained concerning mainly H/C content. Supplemented with XPS, the two spectroscopic tools allowed suggesting reliable hydrogen-and-oxygen enriched atomically-structured models of the predetermined size of the studied ACs basic structural units which were supported quantum chemically.

Amorphous carbons form a large class of carbon allotropes, both natural and synthetic. According to numerous encyclopedic resources, the nomenclature of natural ACs is rather scarce and covers soots and coals, including charcoal. In contrast, the family of synthetic ACs is quite rich including black, glassy, and activated carbons, carbon nanofoams, carbid-derived carbon, carbon fibers, diamond-like amorphous carbon ($\alpha$-C:H and (ta)-C carbons), and others (see review [2] and references therein). The list of synthetic ACs continues to expand since more and more techniques appear to produce this highly requested material. Evidently, it should be complemented with 'technical graphenes' that cover a large class of high-tech new materials produced in the course of reduction of graphene oxide [3]. The list of natural ACs has been unchanged for a long time until scrupulous studies of graphene-like shungite carbon [4-6] and anthraxolite [7] have provided a new pulse which convincingly pointed that these two mineral products should be included as well. Similarly, diamond-like natural ACs have become top issues of the carbon mineralogy of the last two decades [8, 9] as well.

Many of the problems encountered in characterizing the great variety of ACs are due to their massive heterogeneity and a large variation of the main parameters, among which the carbonization rank is the most important. The latter is usually characterized by atomic H/C and O/C contents. The least ratios evidence the highest carbonization and attribution of the relevant amorphous species to the elite group [2]. Samples, selected for the current study, are from this group and are presented by two natural and two synthetic (engineered) $sp^2$ ACs, namely, shungite carbon (ShC) from Shun'ga deposit (Karelia, Russia) and anthraxolite (AntX) from Pavlovsk deposit (Novaya Zemlya, Russia), from one side, and two different Aldrich-Merk carbon black products – No. 699624 (CB624) and No. 699632 (CB632) nominated as "C-99.95%" products [10]. Additionally, high-quality powdered graphites of different origin were used to play the role of reference materials. Thus suggested approach "Selected set of samples + selected set of analytical tools" is aimed at the transformation of 'amorphous' representation of the $sp^2$ amorphous carbons based on particulars, into a 'crystalline' one based on a limited set of fixed commonalities and implies the application of each tool to the whole set of samples.

The paper is organized in the following way. Section 2 presents description of the test samples, following the sequence of methods specified in the introduction. A summarized picture of the structure and chemical composition of the selected samples caused by the presence of oxygen is given in Section 3. Model structures of the basic structure units (BSUs) of the studied ACs, based on the obtained structural-chemical characteristics and supported by QCh consideration, as well as a detailed discussion of common characteristics of model BSUs are discussed in Section 4. Conclusion suggests the main results of the performed study as answers to question what normal, expected, and new was obtained concerning the modern vision of $sp^2$ amorphous carbons.

## 2. Characterization of tested amorphous samples

### 2.1. Sample preliminary treatment

ShC and AntX were produced from the hand-picked samples from the vein of open deposit by grinding to powder with particle size less than 40 μm. ShC was additionally cleaned by repeated several successive treatments in water with stirring and followed by filtration at room temperature. After drying and rubbing homogeneous macroscopic powder with particles less than 5 μm was obtained. CB632 and CB624 were provided by the Sigma-Merk company with the indication of carbon content not less than 99.95 wt% and mean pore diameter of 6.4 nm and 13.7 nm, respectively [10]. All the samples are small-grained, consisting of 10 nm – 10 μ conglomerates. Since ShC and AntX are porous as well [4-7], all the samples may adsorb water due to which four *as prepared* samples (conditionally, *wet* samples) discussed above were complemented by four others subjected to prolong heating (drying) at moderate temperature 110-150$^0$C providing a set of *dry* samples.

### 2.2. Testing the sample structure

*TEM of middle and high resolution.* A lot of TEM and HRTEM images of ShC, AntX and carbon black have been obtained by now (see comprehensive reviews [4, 7, 11, 12] and references therein). Figure 1 presents typical examples, a common 'first-glance' similarity of which related to the three materials is clearly visible. Actually, bands on the top panels from several fractions of a nanometer to several nanometers long are clearly visible in the HRTEM images. These bands are the projections of carbon atom planes oriented almost parallel to the electron beam and present graphene-like *basic structural units* (BSUs) thus evidencing their flat graphene molecular structure. The BSUs are grouped in stacks which aggregate with a lot of free space between them, as seen on the bottom panels thus exhibiting the origin of the species porous structure. Graphene-like BSUs and pores made in due course of the BSUs aggregation are two main motives of the $sp^2$ AC structure. In the current paper the attention will be concentrated on the first issue mainly while a lot of information concerning the porous structure of the species can be found in literature. We will start from the determination of the BSUs structure, a detailed study of which was performed by using X-ray and neutron powered diffraction.

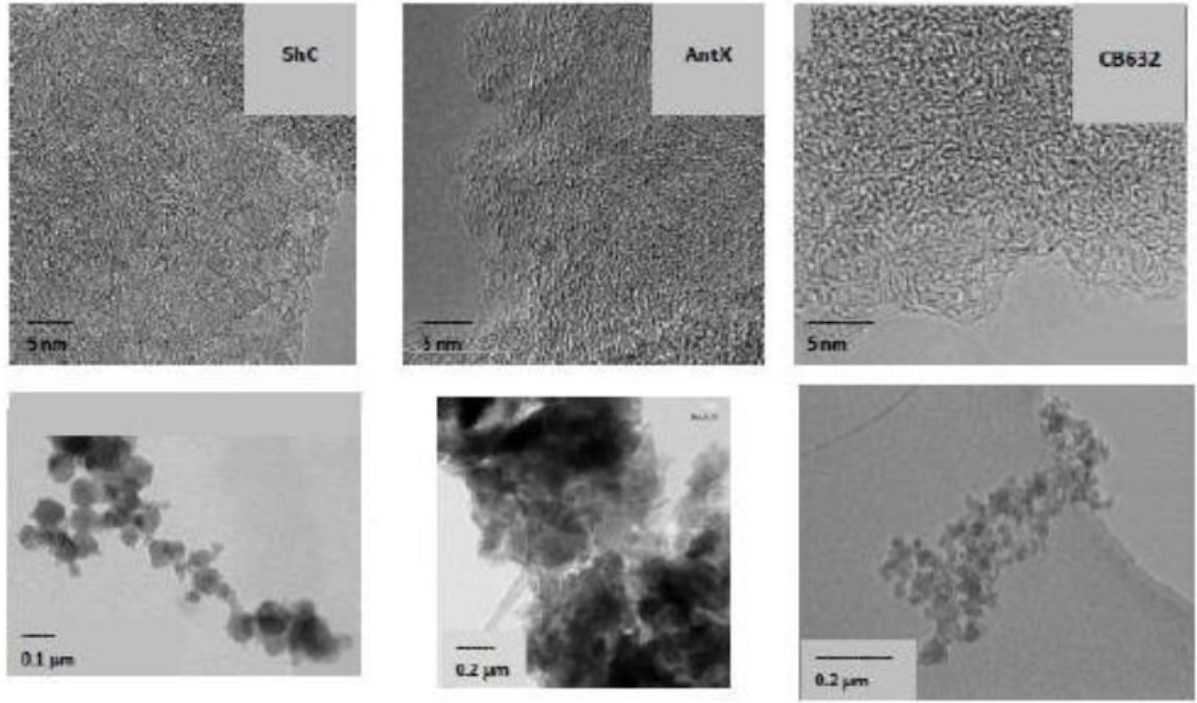

**Figure 1.** HRTEM (top) and TEM (bottom) images of shungite carbon (a), anthraxolite (b), and carbon black (c). Adapted from refs. [5, 7], [11], and [10, 12], respectively.

*NPD and XRPD.* Neutron scattering study, both elastic and inelastic, providing the recording of NPD and INS signals, respectively, was carried out at the high flux pulsed IBR-2 reactor of the Frank Laboratory of Neutron Physics of JINR by using NERA spectrometer [13]. The detailed description of the neutron experiment is given elsewhere [14]. XRPD data were obtained by using a standard Shimadzu XRD-6000 diffractometer.

Figure 2 presents a panoramic view of diffraction data for *as prepared* samples, obtained at low (NPD) and room (XRPD) temperature. Similarly to the reference graphite spectra, given at the bottom, the characteristic features of the ACs diffractograms concern isolated and grouped peaks related to Gr(002), Gr(101) and Gr(110) reflections, the former of which is a dominant. The Gr(002) reflexes are located in the region of 3.3–3.5 Å which determines $d_{002}$ interfacial distance along *c* axis between the neighboring graphene layers in graphite thus pointing to undoubted graphite-like stacking of the relevant BSUs. No less important are reflexes Gr(110) located in the region of 1.1-1.3 Å which characterize the size of BSUs stacks in lateral *a* directions [15]. As seen in the figure, for all the studied ACs Gr(002) and Gr(110) reflexes are shifted (up and down, respectively), as well as considerably broadened pointing convincingly to a considerable space restriction for the relevant BSUs in both directions.

The broadening of Gr(hkl) reflexes is usually attributed to the narrowing of the coherent scattering region (CSR) of a scatterer in the relevant direction. According to Scherrer's formula, the full width at half maximum (FWHM) of diffraction peak *B* and the CSR length $L_{CSR}$ are inversely connected:

$$L_{CSR} = k\lambda / B\cos\Theta \qquad (1)$$

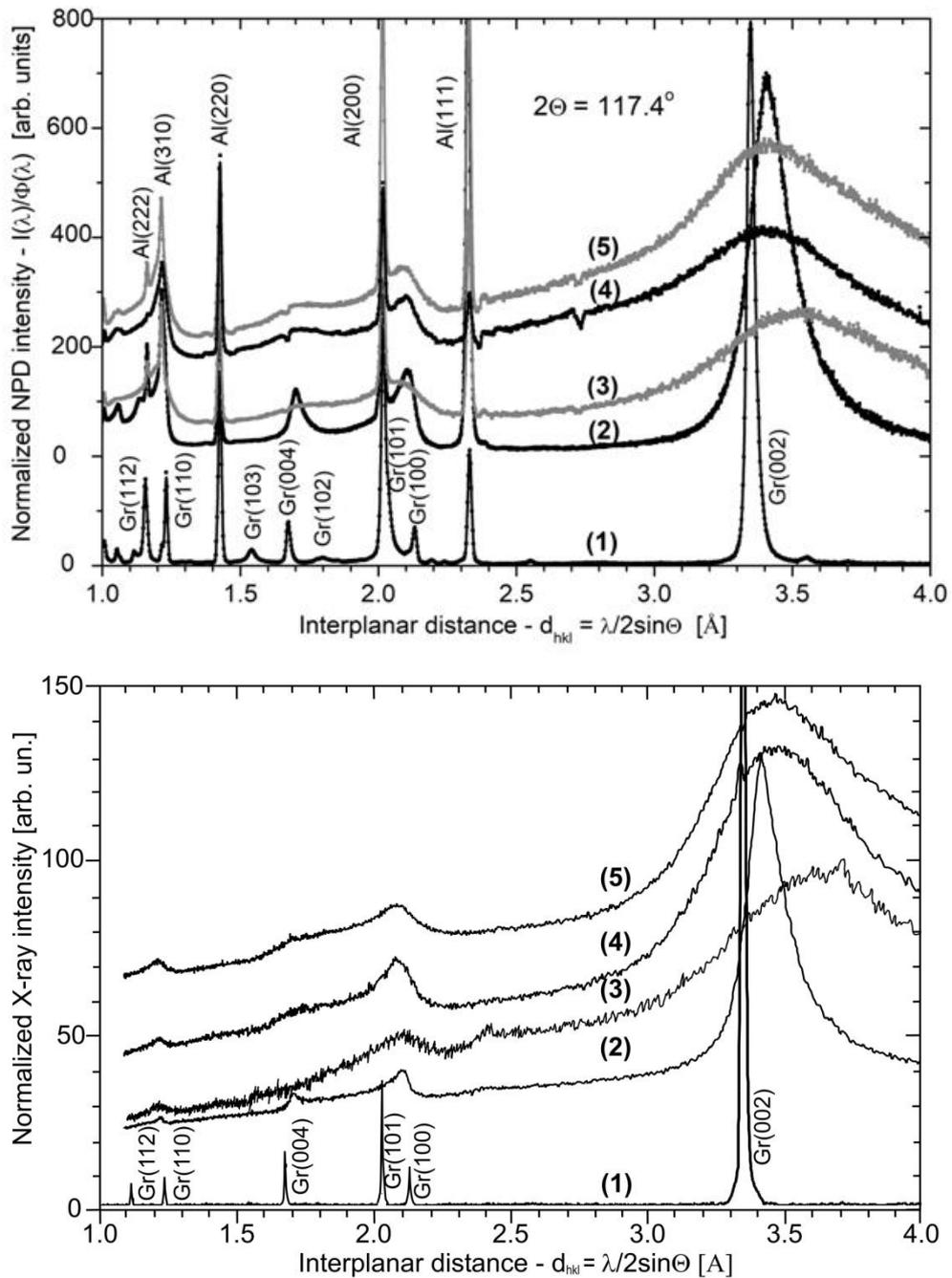

**Figure 2**. Panoramic views of diffractograms of *as prepared* amorphous carbons and graphite (Gr). (a) NPD spectra: 1 – Gr (high purity graphite grade spectral, 20K); 2 – CB624 (6K); 3 – CB632 (6K); 4 – ShC (20K); 5 - AntX (6K), scattering angle $2\Theta=117.4^0$. (b) XRPD spectra at room temperature, 1 – Gr (natural graphite from Botogol'sk deposition), other samples of the same marking.

Here $\lambda$ and $\Theta$ are the neutron and/or X-ray wave length and scattering angle while $k$ is a factor depending on the reflex under study [16]. The factor determination is a permanent problem of a quantitative diffraction study of nanosize objects. However, when the study is performed for a set of samples under the same conditions, it is possible to take one of them (graphite in the current case) as a reference and to determine $L_{CSR}$ of the testing samples addressing to that of the reference. Accordingly, $L_{CSR}$ of the studied ACs are determined as [14]

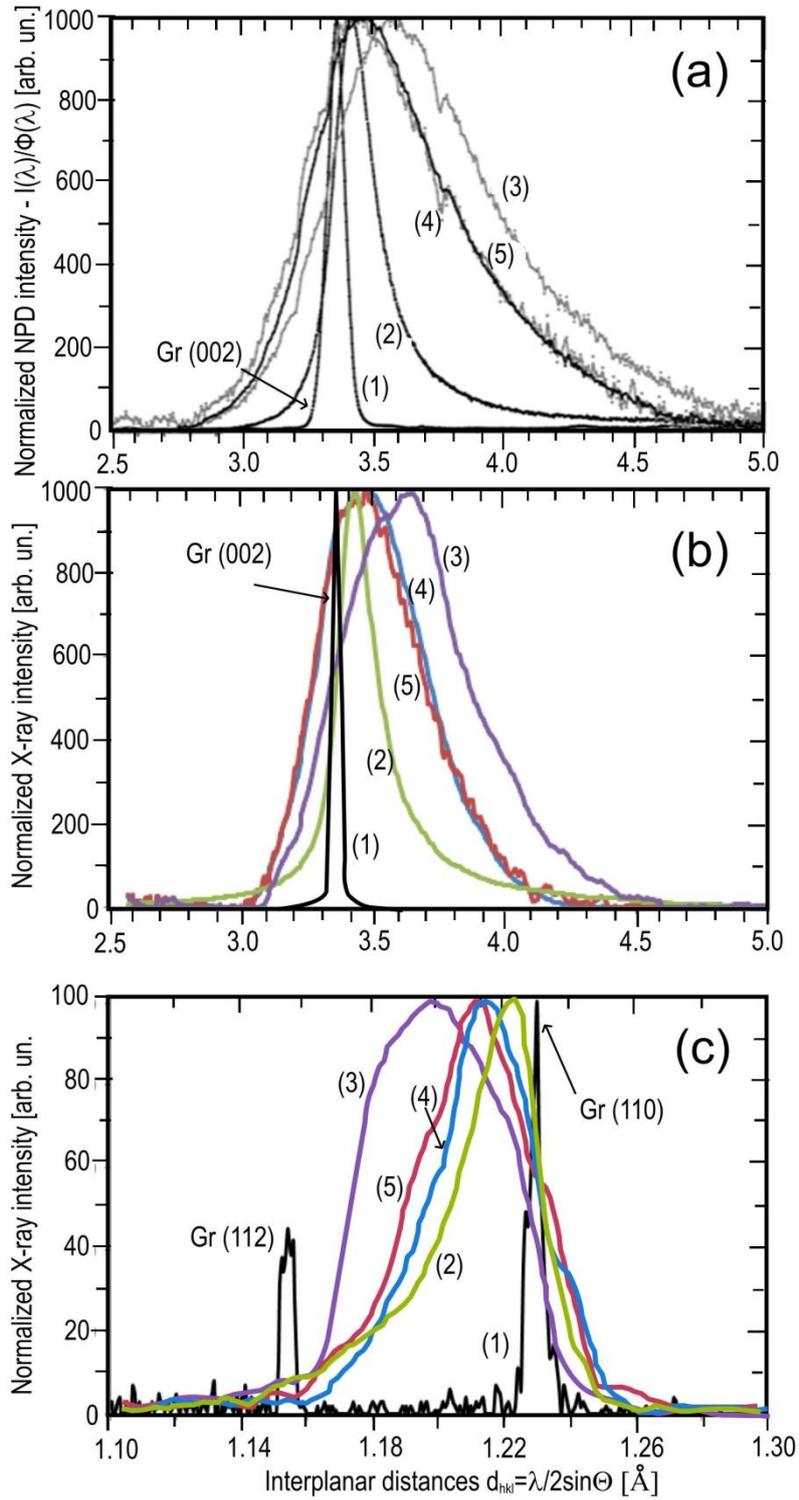

**Figure 3.** Normalized intensities of the NPD (a) and XRD (b) Gr(002) as well as XRPD Gr(110) (c) reflexes after subtraction of linear background of inelastic scattering for the *dry* (NPD) and *as prepared* (XRD) samples. Temperature regime: NPD - 6-20K (see caption to Fig. 2); XRD – room temperature. Sample marking: 1 – Gr; 2 – CB624; 3 – CB632; 4 – ShC; 5 – AntX.

$$L_{CSR} = \left(B_{ref}/B\right)\left(\lambda/\lambda_{ref}\right)L_{CSR}^{ref}. \qquad (2)$$

$L_{CSR}^{ref}$, attributed to crystalline graphite, constitutes ~20 *nm* along both *c* and *a* directions [17].

A comparative view of Gr(002) and Gr(110) reflexes of the studied samples is presented in Fig. 3. To make their analysis more illustrative, all the peaks are equalized by the maximum high. As seen in Fig. 3a and b, NPD and XRPD reflexes Gr(002) are well the same. As occurred, the adsorbed water does not influence markedly both NPD and XRPD plottings that are identical for both *as prepared* (*wet*) and dried (*dry*) samples. Gr(110) reflexes are presented in Fig. 3c for XRPD only, since intense Al(310) reflex of cryostat does not allow a reliable extraction of the peak from the NPD data. The only attempt has been undertaken for ShC [14]. The XRPD reflexes are low-intense and quite noisy due to which they are presented in the figure by trend lines. $L_{CSR}^c$ and $L_{CSR}^a$ data, obtained by applying Eq.(2) to the data shown in Fig.3, are listed in Table 1.

**Table 1.** NPD and XRPD structural data

| Samples | NPD Gr(002) | | XRPD Gr(002) | | XRPD Gr(110) | | Number of BSU layers | |
|---|---|---|---|---|---|---|---|---|
| | $d_{max}$, Å | $L_{CSR}^c$, nm | $d_{max}$, Å | $L_{CSR}^c$, nm | $d_{max}$, Å | $L_{CSR}^a$, nm | NPD | XRPD |
| Graphite[1] | 3.34 | - | 3.35 | - | 1.210 | - | - | - |
| ShC | 3.47 | 2.5 | 3.48 | 2.0 | 1.195 | 2.1 | 7 | 5-6 |
| AntX | 3.47 | 2.5 | 3.47 | 1.9 | 1.191 | 1.6 | 7 | 5-6 |
| CB632 | 3.57 | 2.2 | 3.58 | 1.6 | 1.179 | 1.4 | 6 | 4-5 |
| CB624 | 3.40 | 7.8 | 3.45 | 4.1 | 1.204 | 2.5 | 23 | 12 |

[1] Spectral graphite and natural graphite from Botogol'sk deposition, Buryatia, Russia in the case of NPD and XRPD studies, respectively

As seen in the table, maximum positions, $d_{max}$, of Gr(002) reflexes, provided by NPD and XRPD measurements of the studies ACs, convincingly evidence that all the samples consist of stacks formed by graphene-like BSUs. For ShC, AntX and CB632, the data of the two series are in good consent while remarkably differing for CB624. On the background of grave complexity of the two diffraction events (see, for example, discussions in [15, 18-20]) and high disordering of the sample structure, the first appearance is more surprising than the second. Actually, different character of the diffracting flux interaction with the object (electronic for HRPD and nuclear for NDP), multiplied by the object structural and chemical inhomogeneity and strengthened by different state of the object under experiment (bulk transmission of neutrons through a big sample and depth-limited-area X-ray reflection from quite small species), could provide an expectation of considerable changes between XRPD and NPD patterns. However, such a change is big for CB624 only while the dispersion in maximum positions for the other three ACs can easily be attributed to standard statistical errors as well as to errors caused by the difference in the resolution functions of the neutron and X-ray spectrometers.

The $L_{CSR}^c$ values listed in Table 1 show much more pronounced XRPD/NDP difference, which indicates that the peak broadening is caused by not only size of BSU stacks, but other factors as well, such as inhomogeneous distribution of the stacks over their lateral size and thickness [18-20], the latter due to the difference of both BSU layers dimension and interlayer distance caused by changing chemical framing of the BSUs circumference (chemical gradient

[19]). The obtained data evidently support a common opinion that the application of Scherrer's formula for obtaining size of nanostructure elements from experimental diffraction data has mainly an evaluative character [16, 18-20]. Nevertheless, so far diffraction study has been the only source allowing approaching this type of structure information, the precision and reliability of which should not be overestimated as often is made [21].

Taking these data as real, we can conclude that, basing on NPD data, the thickness of ShC, AntX and CB632 BSU stacks is practically the same and constitutes 2.0 -2.5 nm while that of CB624 is ~3 times bigger. Accordingly, stacks of the first three ACs involve 6-7 BSU layers while stacks of CB624, consisting of 23 layers, may be confidently attributed to nanographite. XRPD data qualitatively support these data suggesting ~25% decreasing in value for the first three samples and ~50% for CB624. So considerable decreasing in the last case is connected with both the Gr(002) reflex narrowing and decreasing $d_{max}$ discussed above.

Embarrassing accessibility of Gr(110) reflexes in NDP and their low intensity in the XRPD plottings makes the $L^a_{CSR}$ determination much more difficult. As seen in Fig. 3c, the relevant XRPD peaks are quite noisy due to which the data listed in Table 1 are not so accurate as for $L^c_{CSR}$ and can be considered as more approximate. Nevertheless, the data clearly reveal that BSUs of all the studied ACs are size restricted in the molecular plane. Evidently, $L^a_{CSR}$ determines the top limit of the BSU lateral dimension while that one may be much less. Actually, once turbostratically arranged, BSUs can simulate much larger lateral size of stacks. On the other hand, the above comparative study basing on NPD and XRPD reflexes showed that the XRPD data may be ~25% underterminated. In view of this, the $L^a_{CSR}$ data given in Table 1 can be considered as a quite reasonable evaluation of the lateral dimensions of the studied BSUs.

Concerning the studied ACs, a comparative analysis of the NPD and XRPD data has allowed revealing that the selected natural and synthetic ACs do belong to graphitic carbons, the main structural fragments of which consist of stacks of graphene-like BSUs. The stack thickness forms a characteristic series BC624 >> SnC ≈ AntX > CB632 which covers the interval from 7.8 (4.1) nm to 2.2 (1.6) nm for NPD (XRPD), respectively. BSUs of all the studied ACs are size restricted in plane. The corresponding size parameter forms a characteristic series BC624 > ShC > AntX > BC632 which covers the interval from to 2.5 nm to 1.4 nm. The difference in the limit values, related to ShC, CB632 and CB624, well correlates with the difference in the pores size of 2-8nm, 6.4 nm, and 13.7 nm, attributed to these amorphics empirically [10, 22]. Once size restricted in the thickness and lateral dimension, BSU stacks lay the foundation of presenting the studied elite ACs as conglomerates of size-varied nanographites.

The presented vision of the studied ACs is not completely exhausted. It is related to flat stacked nanographites, a general HRTEM picture of which is shown in Fig. 1. However, not only flat, but bent BSUs are often observed for the species [11, 12]. The distance between the bent BSUs still remains the same as in the flat structures thus evidencing that BSUs are framed graphene molecules without any addition on their basal planes. A thorough study showed [3] that nanosize free standing graphene molecule can be bent only mechanically or by the presence of external objects such as, say, solid nanoparticles formed by impurities. In the latter case, graphene sheets willingly overlap the bodies simulating their curvature. Since, as we shall see below, nanosized metal particles always present in both natural and engineered ACs, it is quite reasonable to attribute stacks of bent BSUs to the case.

## 2.3. Chemical composition analysis

### 2.3.1. Proximate analysis

*TGA and DSC.* The combustion-in-air thermal gravimetric analysis (DTA) and differential scanning calorimetry (DSC) [23] are widely used for the determination of carbon content of carbonaceous species such as polymers [24], biological macromolecules and extracts [25], carbon containing composite materials (see examples in [26, 27]) as well as 'pure' carbon materials such as coals [28], detonation-synthesized nanodiamonds and accompanying detonation soot [29] and so on. The analysis mainly concerns the *brutto* content of sample carbon, so called *fixed carbon*, since the technique does not distinguish carbon from other combustive ingredients involved in the sample content. Since a complete chemical analysis involves all the ingredients, TGA and DSC are techniques of a proximate analysis.

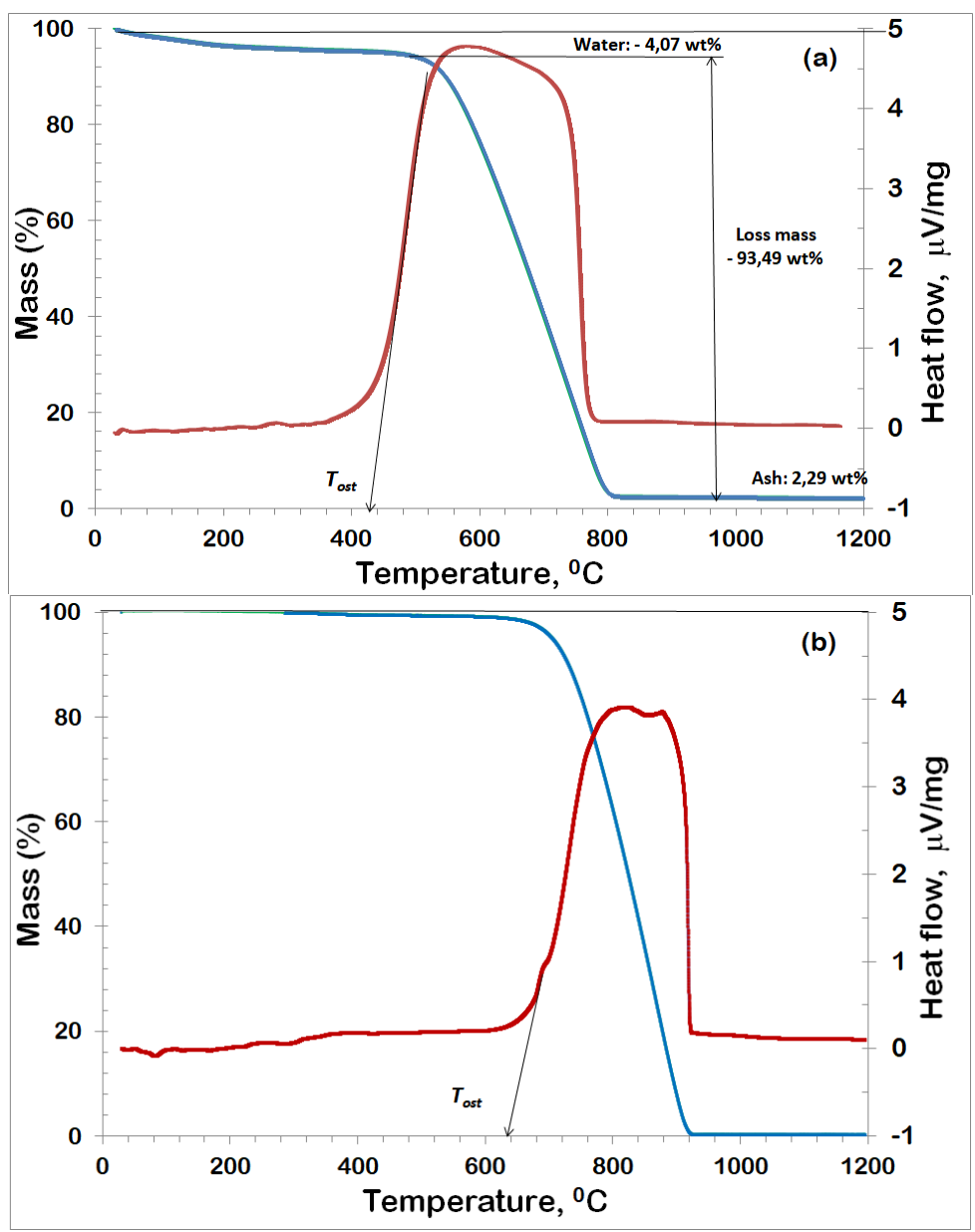

**Figure 4.** TGA thermograms of *as prepared* amorphous carbons ShC (a) and CB624 (b).

In the current study the thermal analysis was conducted on a simultaneous TGA/DSC calorimeter NETZSCH STA 449F1 Jupiter in a dynamic mode at the airflow of 20 ml/sec. The sample was heated from room temperature to 1200°C at a rate of 10°/min with TGA and DSC resolution constituting 0.25 μg and <1 μV, respectively. The obtained thermograms were processed using Proteus software. Samples of ~10mg were carefully weighed and loaded into the instrument's platinum crucible. The change in weight of the sample was automatically recorded until there was no further change. The residue in the crucible represents the ash content of the sample. Simultaneously registered DSC thermograms fixed the heat flow required to increase the sample temperature over the reference one. Figure 4 exhibits a part of the measured data to illustrate the obtained characteristics. As seen in the Fig.4a, ShC heating from room temperature to onset of combustion at $T_{ost}=425^0$C provides releasing of adsorbed water of 4 wt% of the total mass. The combustion proceeds till T=760$^0$ C and provides 97 wt% mass loss. The residual ash is 2 wt% of the total mass. The quantity of fixed carbon of 'pure' carbon materials is usually determined as [28]

$$\% \text{ Fixed Carbon} = 100 - (\% \text{ Adsorbed Water} + \% \text{ Ash} + \% \text{ Volatile Matter}). \qquad (3)$$

In the case of the studied ACs, the amount of volatile matter is evidently negligible due to the absence of characteristic peaks on the DSC thermograms in the temperature region of $250^0$-$400^0$ C that are so pronounced in coals [28]. Accordingly, this ingredient is not taken into account and the amount of fixed carbon is determined by Eq.(3) excluding volatile matter. In spite of a proximate character of the analysis, TGA/DSC thermograms can clearly reveal a qualitative difference between the samples. Actually, once quite similar to all the studied samples, TGA/DSC plottings presented in Fig. 4b evidence that CB624 remarkably differs from ShC. As seen, the CB624 adsorbed water and ash is close to nil. The DSC curve is narrower and the relevant combustion starts at higher temperature $T_{ost}=635^0$C, which tells about different combustion processes occurred for the two ACs at the same conditions.

Thus obtained TGA quantities of fixed carbon, adsorbed water and residual ash related to the studied samples are collected in Table 2. As seen in the table, the contribution of adsorbed water does not exceed a few percents and is the biggest in ShC (4 wt%) while the smallest in CB624 (0.6 wt%). The ash contribution is of the same order and is bigger for natural ACs and less for engineered CBs. The latter is evidently due to intense washing of technical products from ash during the technological cycle. The fixed carbon is definitely not carbon-pure and should include hydrogen and oxygen components at least that are highly combustible constituents besides carbon.

**Table 2.** Proximate mass content of *as prepared* amorphous carbons

| Samples | Fixed carbon, wt% | Adsorbed water, wt% | Ash, wt% |
|---|---|---|---|
| ShC | 93.8 | 4 | 2.2 |
| AntX | 92.3 | 2.4 | 5.3 |
| CB632 | 97.6 | 1.5 | 0.9 |
| CB624 | 99.1 | 0.6 | 0.3 |

If TGA measurements are concentrated on mass loss, DSC plottings exhibit the development of the combustion during the heating. As seen in Fig. 5, the combustion proceeds quite differently when going from ShC to graphite, which can be characterized by changing $T_{ost}$ from 425$^0$C to 720$^0$C, respectively. The modern presentation of the process shows that it is greatly variable depending on the characteristic features of samples. Since all measurements in the current study were performed under the same technical conditions, we should attribute the observed variability to these very features. Among the latter there are such characteristics as sample packing, size of sample structural elements, solubility of evolved gases, heat of reaction, thermal conductivity and so on [23]. Additionally, the features, at least partially, should be connected with the sample porous structure, pores size change of which may cause the $T_{ost}$ depletion [29]. However, as shown, the depletion does not exceed a few centigrades when the pore size is of 10 nm or bigger, which cannot explain so big change in the $T_{ost}$ for the studied ACs. Apparently, the effect should be attributed to changes in the heat of reaction and thermal conductivity of the combustive matter which is evidently deeply connected with the chemical element composition of the samples providing a mixture of different combustive elements. Actually, as seen in Fig. 5, $T_{ost}$ of graphite, which is evidently the purest carbon material and a pretty standard DSC thermogram of which is observed elsewhere [26], is the largest. In contrast, on the opposite side of the observed series of the DSC thermograms, presented in the figure, one can see the plotting related to reduced graphene oxide or technical graphene [3], $T_{ost}$ for which constitutes $370^0C$. The plotting was reconstructed from experimental data presented in [27]. As known, technical graphene, both natural and synthetic, usually contains a few wt percent of hydrogen [30, 31], which is highly favorable for combustion. Accordingly, the presented series of DSC thermograms can be considered as an evidence of a gradual depletion of hydrogen content in the ACs series from ShC to CB624 and graphite. We shall come back to the issue in Section 2.3.3. The tight dependence of $T_{ost}$ on the H/C atom ratio has recently received a strong confirmation by DSC study of carbonized acenes from naphthalene to pentacene exhibiting the $T_{ost}$ gradual increasing from $103.3^0$ C to $419.4^0$ C in the series [32].

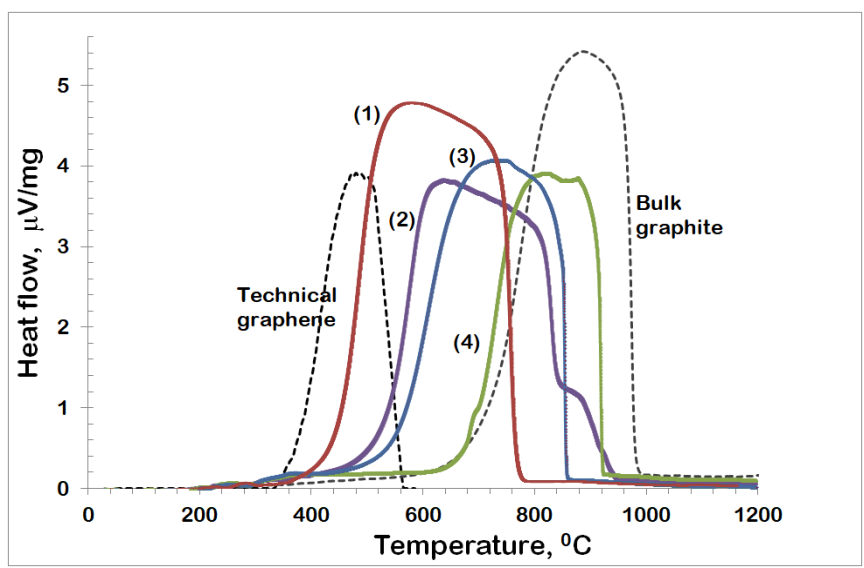

**Figure 5**. DSC thermograms of *as prepared* amorphous carbons: ShC (1), AntX (2), CB632 (3), CB624 (4), and spectral graphite. DSC curve of technical graphene is reconstructed from experimental data presented in Ref. [27].

The raised question about chemical additives in fixed carbon forces to look at the chemical content of ashes of natural ACs as well. The ash residues in the TGA thermographs of Shc and AntX are evidently connected with inclusions into these mineral substances. Actually, Fig. 6 presents the studied natural ACs by EDS map (see Section 2.3.2) related to X-ray emission from basic carbon atoms as well as by the emission of the most contributing silicon and vanadium inclusions. As seen in the figure, the total impurity contribution into AntX is obviously bigger than that of ShC, which might explain the bigger amount of ash at the AntX TGA thermogram. Weight content of these impurities alongside with other ones is presented in Table 4.

### 2.3.2. Ultimate analysis

*Combustion-based elemental analysis (EA).* As clear from the previous section, a proper EA should concern fixed carbon. It means that only water-ash-free samples should be analyzed. In practice, both natural and synthetic samples can be easily released from adsorbed water by a moderate heating while external inclusions, mainly forming the registered ash, still remain. As seen in Table 2, in the current case, the issue concerns natural ACs mainly and this factor should be taken into account when discussing the final tests. The EA was performed for *dry* samples by using combustion-based Vario EL cube Elementar analyzer. Measurements were performed under catalytical tube combustion of samples, separation of foreign gases and separation of the desired measuring components by using thermal conductivity detection. The obtained data are summarized in Table 3. Standard CHNS EA does not determine oxygen content directly and the relevant presented data are just residual content of 100 wt% samples mass after excluding all other contributions. Deviations listed in the table are determined over three independent measurements.

**Table 3**. Ultimate atomic content of *dry* amorphous carbons by CHNS elemental analysis, wt%

| Samples | C | H | N | S | Sum | O residual |
|---|---|---|---|---|---|---|
| ShC | 94.44±0.08 | 0.633±0.023 | 0.88±0.03 | 1.115±0.137 | 97.07 | 2.93 |
| AntX | 94.01±0.35 | 1.105±0.047 | 0.86±0.04 | 1.358±0.066 | 97.34 | 2.66 |
| CB624 | 99.67±0.12 | 0.18±0.026 | - | - | 99.85 | 0.15 |
| CB632 | 97.94±0.03 | 0.316±0.003 | 0.04±0.06 | 0.676±0.003 | 98.98 | 1.02 |

As seen in the table, the carbon content of natural ACs does not exceed 94 wt% while it approaches 98 wt% and 99 wt% for synthetic CB632 and CB624. The former ACs are remarkably enriched by hydrogen, nitrogen, sulfur, and oxygen. For synthetic ACs, only for CB632 the contribution of other elements is significant. Therewith, that one of sulfur and oxygen approaches 1 wt% while the hydrogen content is three times less. The presented data are related to the samples volume as a whole.

*Energy-dispersive spectroscopic analysis (EDS).* The technique is provided by scanning electron microscopy (SEM) complemented with spectroscopic mapping by energy dispersive X-rays (EDX) (SEM-EDX) or energy dispersive spectroscopy (EDS) (SEM-EDS) [33]. The current measurements were carried out by using X-MAX detector of Oxford Instruments coupled to the Tescan MIRA3 LMH electron microscope. The performed analysis was both qualitative and quantitative. The former, known as EDS mapping, allows registering the distribution of a selected element over the sample by monitoring a monochromatic X-ray emission characteristic for the element. Such EDS maps are presented in Fig. 6 for ShC and AntX, exhibiting carbon, silicon and vanadium atoms distribution over the samples. The quantitative EDS analysis concerns a comparison of the intensity of element-wave-dependent X-ray emission with that one of a standard element reference registered at identical conditions.

In the first approximation, the X-ray emission intensity related to a characteristic line $I_{at}$ of a selected element is proportional to the element concentration $C_{at}$. The linear dependence is valid for both the studied element and the standard. Therefore, the sought-for weight of the atom fraction $C_{at}$ in the sample can be estimated by a comparison with that one in the standard $C_{st}$ following the relation

$$C_{at} = C_{st} \cdot (I_{at}/I_{st}) \qquad (4)$$

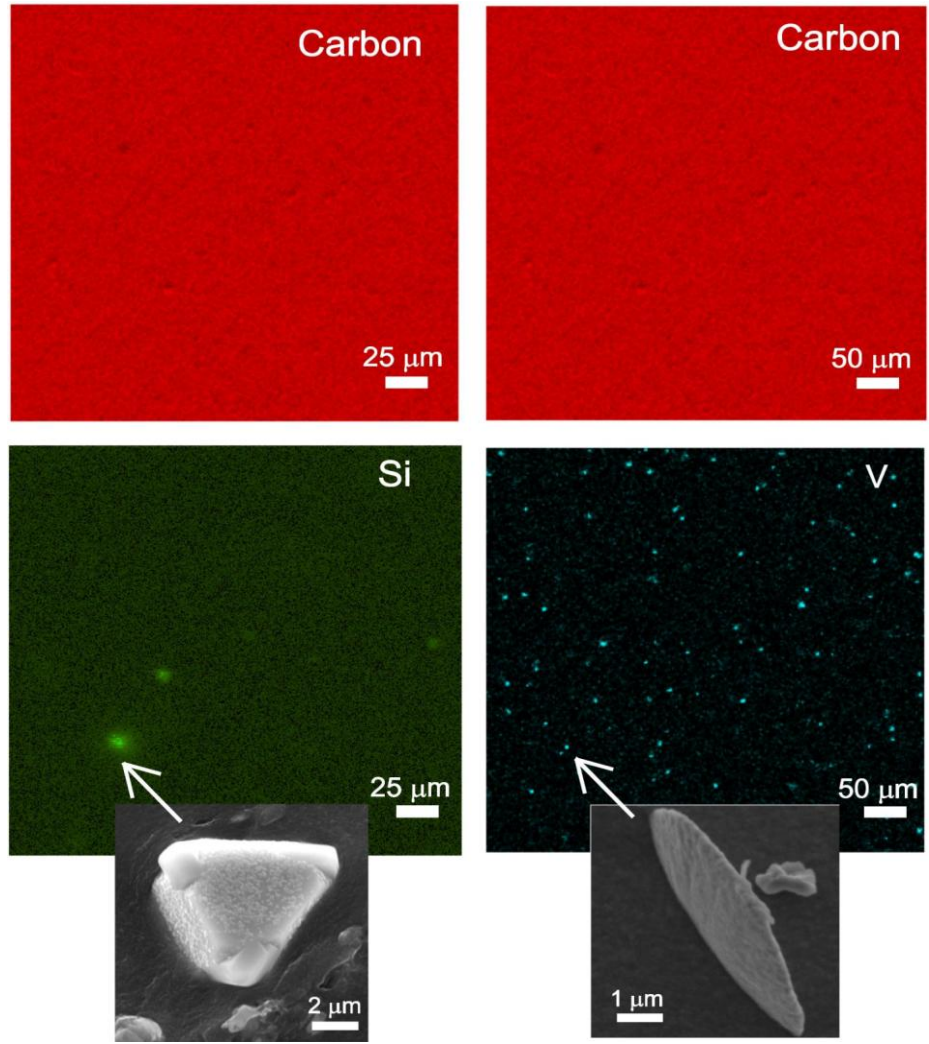

**Figure 6**. Typical EDS maps of carbon for ShC (top left) and AntX (top right). Distribution of silicon in ShC (left middle) and vanadium in AntX (right middle). SEM images of mineral inclusions indicated by arrows (bottom).

In the current study, C- and O-atom fractions in the studied ACs are determined with respect to rod glassy carbon (Aldrich-Merk) and $SiO_2$ as standards, by which the X-MAX detector of Oxford Instruments is equipped. C-standard is usually considered as monoelemental substance with $C_{st}=1$. However, since the standard belongs to the $sp^2$ ACs similar to the studied CB632 and CB624, which occurred to be not monoelemental as shown earlier, the sample was involved in the EDS analysis as well. Actually, the resulted C-atom fraction $C_{at}$ constitutes 0.97. Other weight % contributions related to the standard are given in Table 4. As seen in the table, the summary weight percentage constitutes 100.69 wt%. The total deviation from 100 wt% value less than ±1 wt% is usually accepted as the evidence of satisfactory self-consistent analysis.

Figure 7 presents a collection of EDS spectra of the studied *dry* samples in the energy region up to 6 keV while partitioned chemical components are presented by spherical diagrams. The relevant quantitative wt% contents are given in Table 4. The measurements were performed at inclusion-free sites controlled by EDS maps similar to Si- and V-maps of ShC and AntX presented in Fig. 6. The data were averaged over those ones determined at five different places for each samples. As seen in the figure, the emission spectra are presented by main signals related to carbon, amount of which covers the region of 95.05-97.44 wt%, and oxygen of 1.7-3.3 wt%. Besides, a variable set of minor impurities is seen for different samples. The latter are the most numerous for ShC and CB632 while the scarcest in the case of AntX and CB624. As seen in the table, the summary percentage of all the samples is close to 100% evidencing a convincing reliability of the results obtained. EDS analysis confirms that the studied ACs are not carbon-pure but involve additionally a few percent of oxygen and ~1.5 wt% of minor impurities. The technique is not sensitive to hydrogen.

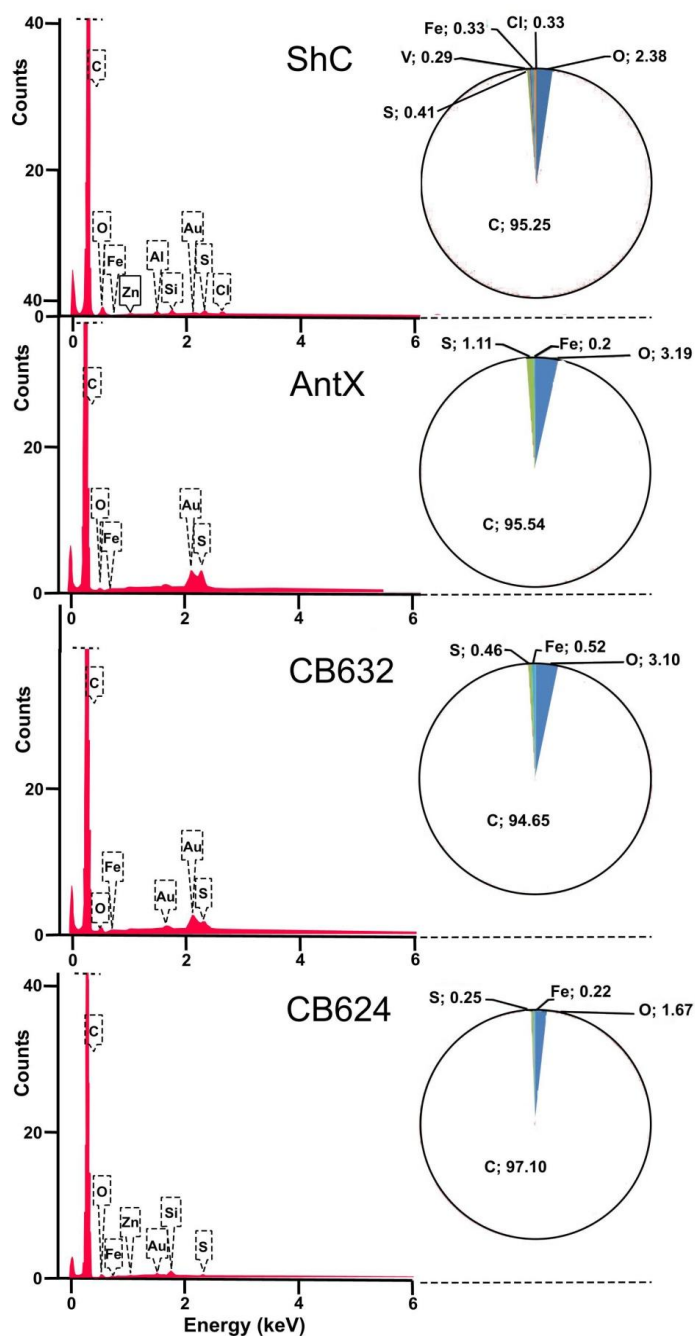

**Figure 7.** EDS spectra and content diagrams of *dry* amorphous carbons. The spectra were recorded on a Gatan-797 slow scan CCD camera. Spherical diagrams visualize the samples chemical content, details of which are presented in Table 4.

Table 4. EDS ultimate atomic content of *dry* amorphous carbons, wt%

| Samples | Elemental content | | | |
|---|---|---|---|---|
| | C | O | Minor impurities | Summary percentage |
| C-standard | 97.05 | 1.87 | **S** - 1.11; **Zn** – 0.66; **Fe** – 0.19 | 100.69 |
| ShC | 95.16 | 3.31 | **S** - 0.41; **Zn** - 0.84; **Fe** - 0.33; **Cl** - 0.33; **Na** - 0.13; **K** - 0.05 | 100.56 |
| AntX | 95.54 | 3.19 | **S** - 1.11; **Zn** - 0.66; **Fe** - 0.19 | 100.69 |
| CB632 | 95.05 | 3.1 | **S** - 0.47; **Zn** - 1.06; **Fe** – 0.52; **Cl** – 0.07; **Al** – 0.07; **Si** – 0.09 | 100.44 |
| CB624 | 97.44 | 1.68 | **Zn** - 0.6; **Fe** - 0.22 | 100.32 |
| Graphite Botogol'sk deposition | 99.17 | 0.42 | **Fe** – 0.34; **Cr** – 0.09 | 100.02 |

*XPE survey spectra.* The last contribution to the chemical content determination of the studied ACs was provided by XPS. A general panorama of the survey spectra of X-ray photoemission (XPE) is presented in Fig. 8. XPS was conducted at Specs PHOIBOS 150 MCD spectrometer using X-ray gun of Mg (hv = 1253,6 eV). Vacuum in the chamber does not exceed $3 \times 10^{-9}$ Torr. Samples were placed on conductive tapes (3 MX-7001). High degree of vacuum provided the releasing of the studied samples from adsorbed water due to which the spectra of *as prepared* and *dry* samples were identical. During the further treatment, the background was subtracted basing on Shirley method [34]. Quantification of atomic content was provided using sensitivity factors provided by the elemental library of CasaXPS. The modified asymmetric Voigt type lineshape [35] was used to separate the waveform of XPE spectra. High quality graphite GSM2 (ГСМ2) [36] was used as a reference sample. The XPE spectra of the studied samples look quite similar evidencing contribution of C 1s and O 1s spectra of the main chemical components and allow evaluating atomic percentage of the observed elements by standard technique. The obtained data are given in Table 5. As seen in the table, all the ACs involve a considerable oxygen content. Recalculated data in terms of wt% are listed in Table 6.

Table 5. XPS ultimate atomic content of *dry* amorphous carbons, at%

| Samples | Elemental content | | |
|---|---|---|---|
| | C | O | Minor impurities |
| ShC | 92.05 | 6.73 | **S** - 0.92; **Si** – 0.20; **N**- 0.10 |
| AntX | 92.83 | 6.0 | **S** - 0.85; **Si** – 0.25; **N**- 0.07 |
| CB632 | 93.32 | 6.02 | **Si** – 0.66 |
| CB624 | 95.01 | 4.52 | **Si** – 0.46 |
| Graphite GSM2 | 96.64 | 2.53 | **Si** – 0.83 |

### 2.3.3. Comparative analysis and experimental support

Table 6 summarizes the data related to the chemical content of the studied AC samples by using the above analytical techniques. All the methods used allow direct determining of carbon content. However, the dispersion of the latter occurs quite large and constitutes 95.2-88.5 wt % for ShC, 95.6-89.5 wt% for AntX, 97.94-90.7 wt% and 99.67-93.1 wt% for CB632 and CB624, respectively. The feature clearly evidences multielemental character of the species, on the one hand, and different sensitivity of the used analytical techniques to chemical elements that accompany carbon in the studied samples, on the other. The carbon content is bigger for carbon black products, the biggest for CB624, and expectedly the least for natural amorphics that were originated in much richer chemical surrounding. Important to note that all the reference carbons used in the study, among which there were spectral graphite (NPD, INS, DSK), natural graphite (Botogol'sk deposition, Buryatia, Russia), graphite GMS2 (XPS) and 'C standard' rod glassy carbon (EDS), occurred not to be pure carbons as well (see Tables 4 and 5). Oxygen is the main impurity, which may point a heightened propensity of pure carbons to oxygenation. The hydrogen content is reliably fixed in ShC, AntX, and CB632, but is at the level of statistical nil in the case of CB624 and graphite GMS2.

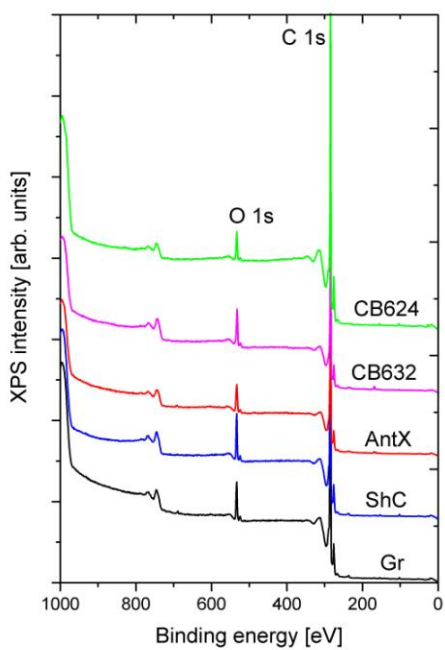

**Figure 8**. XPE survey spectra of *as prepared* amorphous carbons and graphite GSM2 at room temperature.

The next comment concerns the data dispersion. As seen in the table, the later related to first three techniques (TGA, EA HCNS and EDS) is evidently less than that involving XPS data as well. Therewith, if the dispersion difference for carbon content is of a few percents, that one for oxygen reaches a few times. As known, such a significant difference concerning chemical content is characteristic for XPS and is usually attributed to the surface-area character of the latter in contrast to bulk one due to X-ray photoemission occurring from small depth of ~ 2 Å. Another reason concerning the current study will be considered later.

**Table 6**. Summarized ultimate atomic content of *dry* amorphous carbons, wt%

| Elements | TGA[a] | EA HCNS | EDS | XPS |
|---|---|---|---|---|
| **Shungite carbon** | | | | |
| C | 93.8 | 94.44 | 95.2 | 88.5 |
| H | - | 0.63 | - | - |
| O | - | 2.93 | 3.3 | 8.6 |
| Impurities | **w**-4; **ash**-2.2 | N- 0.88; S-1.12 | S - 0.1; Zn - 0.8; Fe - 0.3; Cl - 0.3; Na - 0.1; K - 0.1 | S-2.3; Si-0.5 N-0.1 |
| **Antraxolite** | | | | |
| C | 92.3 | 94.03 | 95.6 | 89.5 |
| H | - | 1.13 | - | - |
| O | - | 2.64 | 3.2 | 7.7 |
| Impurities | **w**-2.4; **ash**-5.3 | N-0.86; S-1.36 | S - 1.1; Zn - 0.7; Fe - 0.2 | S-2.2; Si-0.5; N-0.1 |
| **Carbon black** | | | | |
| **CB632** | | | | |
| C | 97.6 | 97.94 | 95.0 | 90.7 |
| H | - | 0.32 | - | - |
| O | - | 1.02 | 3.0 | 7.8 |
| Impurities | **w**-1.5; **ash**-0.9 | N-0.04; S-0.68 | S - 0.5; Zn - 1.1; Fe – 0.5; Cl – 0.1; Al – 0.1; Si – 0.1 | Si-1.5 |
| **CB624** | | | | |
| C | 99.1 | 99,67 | 97.4 | 93.1 |
| H | - | 0.18 | - | - |
| O | - | 0.02 | 1.7 | 5.9 |
| Impurities | **w**-0.6; **ash**-0.3 | S-0.13 | Zn – 0.6; Fe - 0.2 | Si-1.0 |
| **Graphite** | | | | |
| C | - | 99.69[b] | 99.17[c] | 95.6[b] |
| H | - | 0.12 | - | - |
| O | - | 0.01 | 0.42 | 3.4 |
| Impurities | | | Fe – 0.34; Cr – 0.09 | Si – 1.0 |

[a] Fixed carbon of *as prepared* samples
[b] Graphite GSM2
[c] Graphite of Botogol'sk deposition

The performed analytical study has shown that all the studied ACs are multielemental compounds with dominating contribution of carbon while completed by a set of other minorities. Among the latter, oxygen evidently plays the main role constituting a few wt% in all samples. Hydrogen is the next by importance for carbon materials with graphene-like structure. However, as seen in the table, the hydrogen weight content is comparable with that of other minor impurities involving sulfur, nitrogen, chlorine, silicon, and different metals, role of which is very important in geochemistry of carbon [37, 38] or engineering technology of highly carbonized products [39]. See as well a comment concerning the origin of bent BSUs in Section 2.2. Certainly, the first three elements alongside with oxygen and hydrogen are involved in the formation of the AC BSUs. However, once concentrated on disclosing the BSU molecular structure, we shall restrict ourselves in what follows by the consideration of carbon-hydrogen-oxygen triad only. Evidently, both minor chemical entities are chemically bound with the BSU carbon core. Since, as shown in Section 2.2, both flat and bent BSUs are arranged in stacks with interlayer distance close to the graphite one, neither hydrogen nor oxygen atoms are located on the BSU basal plane but are placed in the BSU circumference area. Therefore, BSUs present nanosize framed molecules of graphene oxyhydrides. The best way to support the triad data content of these molecules is to address detailed spectroscopic studies provided by INS and XPS.

*INS spectra.* INS is one of the most powerful tools to reliably fix the presence of hydrogen in any matter. However, this high-flux-reactor-based neutron source technique does not belong to analytical ones due to which the hydrogen monitoring is usually mainly qualitative. However, a comparative study of a set of selected samples under the same conditions as well as special measures undertaken at the obtained results treatment for a correct normalization of the INS spectra intensity per sample mass to be provided, make the INS study rather quantitative as well. Such a study was performed for *wet* and *dry* samples of the studied ACs [40] and the obtained results give a perfect chance to check data listed in Table 6.

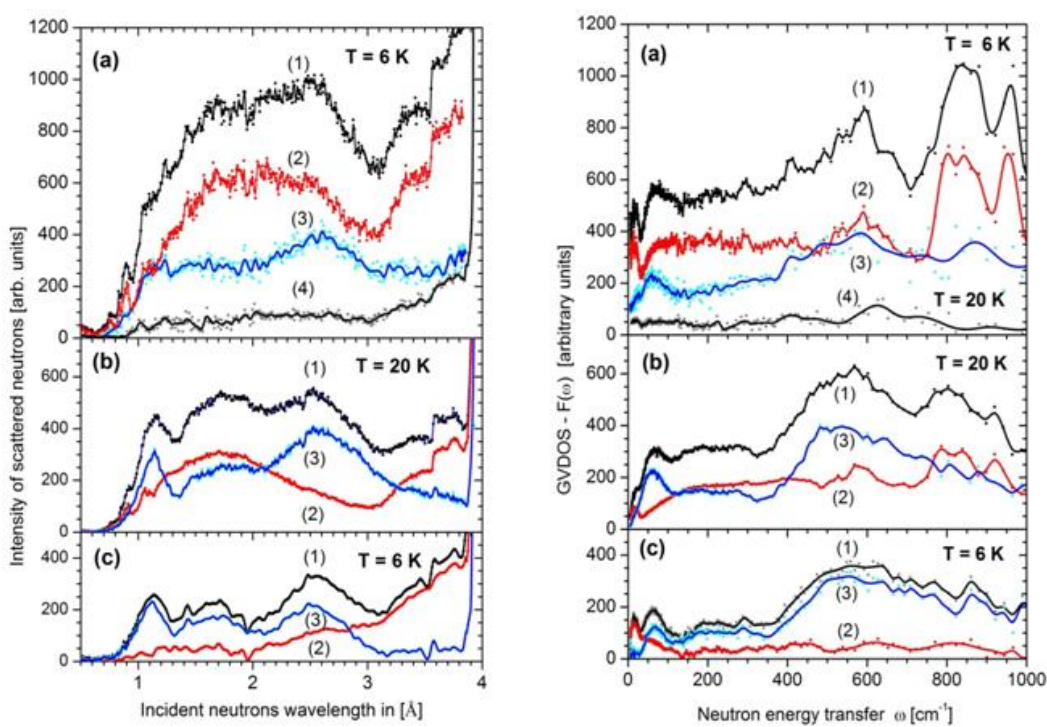

**Figure 9**. Normalized TOF INS spectra (left) and generalized vibration density of states distribution (right) of AntX (a), ShC (b), and CB632 (c). Figures mark spectra of *as prepared* (1) and *dry* (2) samples as well as the differences (3) between spectra (1) and (2) presenting adsorbed water. Curves (4) in panels *a* present spectra of *as prepared* CB624. Adapted from [40].

The left part of Fig. 9 presents a consolidated collection of the obtained INS time-of-flight (TOF) spectra. The spectra are thoroughly treated and normalized per 10g of mass to be able to be compared quantitatively. Black curves in the figure (Spectra 1) are related to *wet* as prepared samples and are resulted from scattering on both adsorbed water and hydrogen-enriched carbon core of the samples. Red curves (Spectra 2) describe the INS spectra of *dry* samples after complete removing of the adsorbed water in due course of a prolong heating at $120^0$ C. Evidently, blue curves (Spectra 3), which are differences between Spectra 1 and Spectra 2, correspond to INS spectra of adsorbed water. The right part of Fig.9 presents the energy-converted INS TOF spectra in terms of generalized vibrational density of states (GVDOS) (see [30] for details). Numbering and color marking of GVDOS spectra is the same as for the pristine INS TOF ones. A comparative analysis within families of Spectra 1, 2, and 3 allows obtaining important information concerning the hydrogen component of atomic structure of BSUs of the studied ACs.

As seen in Fig. 9, the comparison of Spectra 1 convincingly shows that three of the *wet* studied samples, namely AntX, ShC, and CB632, are hydrogen-rich enough while the amount of hydrogen in CB624 is at the level of statistical errors due to which its Spectra 4 in panels *a* cannot be subtracted from the reference spectra of graphite [40]. The feature evidences that the hydrogen amount in the sample is close to statistical nil. This result well correlates with the EA data for CB624 and GSM2 presented in Table 6. A comparing analysis of Spectra 3 shows their good consent with the TGA data related to adsorbed water. Actually, as seen in Fig. 9, the water amount is the least for CB632, while about twice and three times bigger in ShC and AntX, respectively. As seen in Table 6, this serial water amount perfectly correlates with the TGA data. Spectra 2 of *dry* samples reveal the hydrogen content of the carbon core of the studied samples that evidently increases when going from CB632 to AntX. Actually, Spectra 2 of CB632 do not differ much from the spectra of the reference graphite thus pointing to a small contribution of hydrogen albeit the presence of hydrogen in the sample carbon core leaves no doubts. In contrast, the hydrogen availability in the carbon core of ShC was convincingly fixed by INS previously [12, 28, 29] due to which the marked exceeding of the ShC scattering intensity over that one of CB632 is quite expected. AntX completes the hydrogen hierarchy, demonstrating about twofold increase in the scattering intensity in comparison with ShC. The observed semi quantitative correlation of the hydrogen amount in the samples perfectly correlates with the EA data listed in Table 6. Besides exhibiting the hydrogen atoms presence and their amount, INS allows revealing the manner of chemical bonding of the atoms with the carbon core. A convincing conclusion about involving hydrogen atoms in the formation of C-H bonds was made [40]. Therefore, INS study has provided a clear vision of structural aspect of hydrogen atoms in the studied ACs.

*XPE spectra*. In contrast to INS, which is a "hydrogen tool" and which provides the test of H/C content ratio, XPS in the case of carbonaceous materials is an "oxygen tool" and is mainly attributed to O/C content distributed over a number of oxygen-containing groups (OCGs). As in the case of INS, it is highly desirable to get information concerning both the O/C content ratio and the type of OCGs that realize the chemical bonding of oxygen atoms with carbon core. However, the situation occurred more difficult. As was shown in the previous section, XPE survey spectra would have coped quite well with the first task if not for the above-mentioned annoying circumstance concerning the overestimation of oxygen content in comparison with other techniques. Obviously, the feature is connected with the small escape depth of X-rays, which limits the studied volume of samples to a few tenths nanometer in thickness. Looking at Fig. 1, it is possible to imagine that the depth area, commonly attributed to particular 'surface area', covers not the entire BSU molecules, whose size is about an order of magnitude bigger than the escape depth, but only part of them, the size of which depends on the tilting angle of the molecules with respect to the working surface. HRTEM images in the figure clearly show that a great deal of BSUs is oriented practically normal to the surface due to which X-rays see only molecules edges but not entire molecules. Since oxygen atoms must be concentrated on the edges, their content might be evidently overestimated. Therefore, a drastic spatial anisotropy of BSU molecule structure may be responsible for the observed effect.

The analysis of chemical bonding by using XPS is based on two fundamental facts: I) the characteristic value of binding energy (BE) of electrons on internal orbits of atoms $i$, which determined a particular atom-related wave length, $\lambda_{iA}$, of the emitted X-rays and 2) the $\lambda_{iA}$ value dependence on the atom neighboring, which allows disclosing the type of chemical bonding of the atom to the other ones. Both features are well presented in the case of ACs by carbon (C 1s) and oxygen (O 1s) XPE spectra that are shown in Fig. 10 a and b. C 1s spectra are normalized by maximum intensities, while O 1s spectra in Fig. 10b were additionally normalized to be consistent with the oxygen content of the studied samples listed in Table 6. As seen in Fig. 10, both sets of spectra are markedly variable, thus demonstrating varying chemical surrounding of carbon and oxygen atoms in the studied ACs and convincingly evidencing different 'oxygen chemistry' that accompanied the studied amorphics formation.

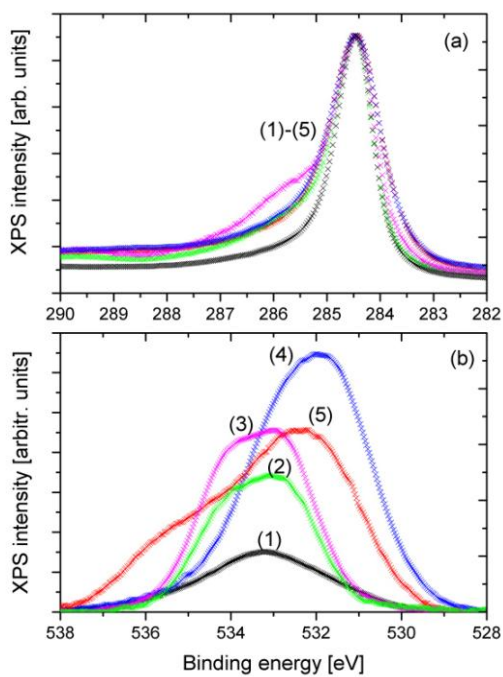

**Figure 10.** C 1s (a) and O 1s (b) XPE spectra of graphite GMS2 (1); CB624 (2); CB632 (3); ShC (4) and AntX (5) at room temperature.

## 3. Oxygen-containing atomic compositions of $sp^2$ amorphous carbon basic structural units

Despite the presence of distinct left-side tail of the main peak of the C 1s spectrum of graphite in Fig. 10a as well as of O 1s spectrum in Fig. 10b, which evidence partial oxygenation of graphite, the peak exhibits the resolution ability of the apparatus quite reliably since the effect of oxygen atoms on the peak width is quite small [41]. C 1s peak of CB624 is practically the same as in graphite while those of CB632, AntX and ShC are clearly broadened. In contrast to oxygen, hydrogen atoms cause a marked effect of the C 1s peak width related to $sp^2$C-H atoms directly connected with hydrogens [42, 43]. Therefore, increasing broadening of the central C

1s peak when going from CB632 to ShG and AntX evidences growing hydrogen-component contribution to the framing area of the relevant BSU molecules. The feature well correlates with characteristics of these amorphics combustion presented in Fig. 5 as well with INS spectra shown in Fig. 9. At the same time, a considerable tailing of C 1s spectra in the left-hand side and pronounced O 1s spectra disclose oxygen component of the BSU molecular framing.

The tailing of C 1s spectra and broad O 1s ones are commonly associated with a multi-variable contribution of carbon and oxygen atoms differently subjected to surrounding. In case of carbon materials, particularly, those related to nanostructured graphene-based ones, until recently the relevant XPE spectra have been analyzed in terms of the 'four-peaks' approximation that involves groups C-C, C-O, C=O, and COO of C 1S spectra and C-O, C=O, C(=O)O, and O=C(O) groups of O 1s spectra [43-55]. In the latter case, the corresponding components are usually associated with the simplest OCGs such as hydroxyls, epoxides, carbonyls and carboxyls. The approach was resulted from the extended XPS study of numerous polymers of different structure which laid the foundation of atomic group assignment to characteristic XPS peaks [42]. However, as shown by extended IR absorption studies, in the case of graphene-like materials, much more complicated OCGs including benzenoid heterocycles such as ketones and quinones, cyclic ester, lactones and acid anhydrides, furan and pyrans, as well as hydroxy pyrans and so on (see for examples [56-58]) are found in oxidative products. Similar questions were raised in the case of XPS [41] as well, a profound review of which was presented by Yamada et al. [59]. Basing on recent studies as well as on extended investigations of chemical properties of black carbons [60-62] and accepting a large complexity of the oxidative products originated in due course of graphene molecule oxidation, the authors performed a profound computational consideration of a large number of model graphene molecules of varied OCG composition. The obtained results were applied to successfully analyze XPE spectra of heat-treated graphene oxide. A sparkling variety of XPE spectra of the studied ACs presented in Fig. 10 makes to think that the feature has a direct relation to the variety of oxidative products and is yet another direct proof of this fact. In light of this, it seems quite reasonable to follow the Yamada et. al. concept, which can be called as 'multi-complex-OCG' (MC-OCG) one, to analyze the obtained XPE spectra of the studied ACs.

Figures 11a and 11b show pristine C 1s and O 1s spectra of the studied samples (black curves) alongside with resulted envelopes (dotted curves) of series of Voigt-fitting-function (VFF) peaks presenting the asymptotic expansion of the pristine spectra. To simplify further comparison, the pristine spectra of both groups, presented in Fig. 10, are normalized per peak maxima. Both C 1s and O 1s spectra are expanded by using 7 and 5 VFFs for C- and O-spectra, respectively. Positions of the input VFF peaks, their assignment with respect to chemical structure as well as resulted output VFF peaks, accompanied with peak positions (PP), FWHM (FW) and partial contribution (PC) into the spectra total intensity (so called PP-FW-PC format) are listed in Table 7.

*C 1s spectra.* The spectra are formed around Peak 2 positioned at 284.3 eV that corresponds to the binding energy (BE) of carbon atoms forming $sp^2$ C=C bonds of unperturbed graphene molecules. According to MC-OCG concept [59], BE is negatively shifted when the atoms are involved in the formation of $sp^2$C-H bonds, which is described by Peak 1 at 283.6 eV. Peak 3 at 285 eV reflects BE shift caused by the presence of other chemicals in the vicinity of carbon atoms while Peak 4 at 285.6 eV corresponds to carbon atoms involved in the formation of C=O bonds. Peaks 5-7 present the main part of the MC-OCG concept involving, beside hydroxyls $sp^2$**C**-OH (Peak 5 at 286.5 eV) and carboxyls O=**C**-OH (Peak 6 at 288.3 eV), O=C-O-**C** and O=**C**-O-C groups of lactones, O=**C**-O-**C**=O of acid anhydride, **C**-O-**C** of cyclic ester and other compositions listed in the table. The entire complex of the groups takes into account practically all set of possible carbon atom perturbations that might occur at hydrogenation and oxygenation of $sp^2$ carbons in general [60-62] and a graphene molecule in particular.

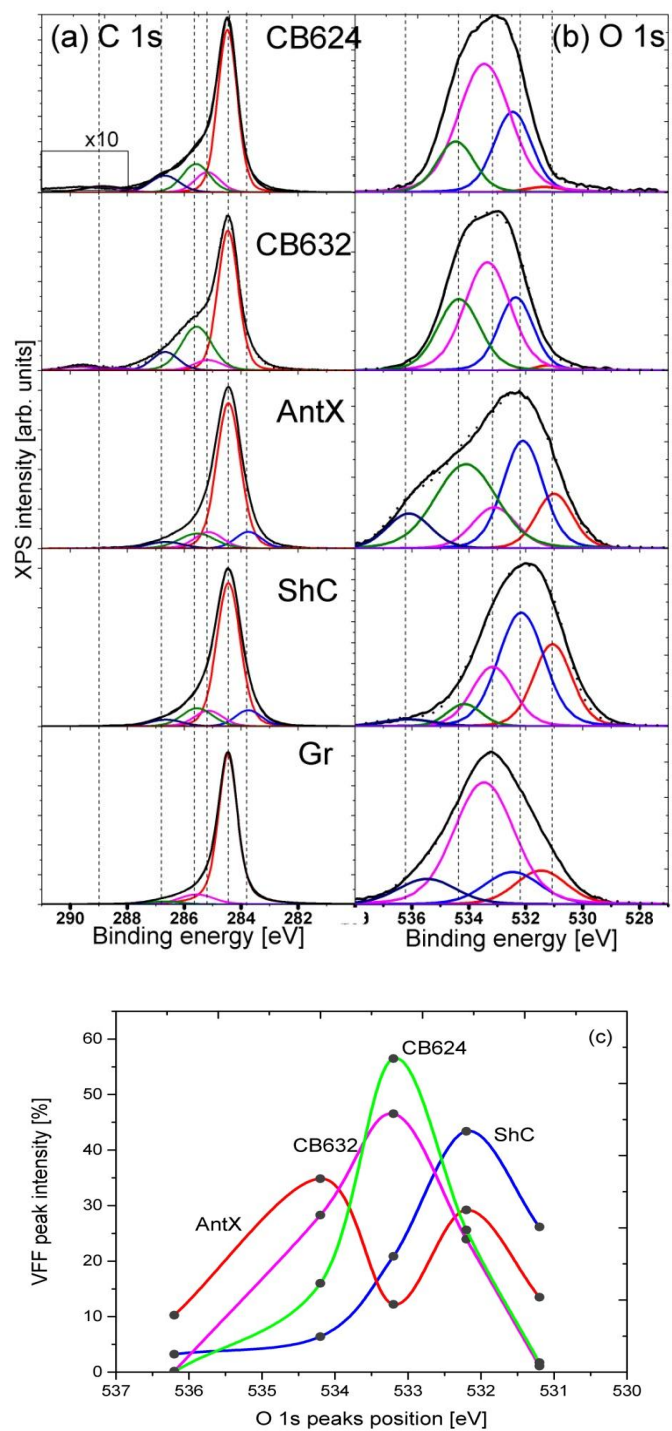

**Figure 11.** Expanded C 1s (a) and O 1a (b) XPE spectra of graphite GMS2 (1); CB624 (2); CB632 (3); ShC (4) and AntX (5) at room temperature. c. Distribution of Voigt-fitting-function peaks of O 1s spectra over peaks number.

The results of the studied C 1s spectra expansion over seven VFF peaks discussed above are listed in the corresponding rows of Table 7. The expansion qualitative verification can be performed by comparing FWHMs of the VFF peaks. The data related to central Peaks 2 may be considered as guiding points. As seen in the table, FWHM of the peak of graphite constitutes 0.8 eV, which together with the absence of Peak 1 in the expanded spectrum points to the attribution of the width to the resolution ability of spectrometer. Basing on the fact, it becomes evident that the width of the VFF peaks cannot be narrower than 0.8 eV. Accepting the feature, much narrower Peak 3 of graphite as well as similar Peaks 1 of CB624 and CB632 were excluded from the input and output data. As for FWHM data for all other peaks, those fill the region of 1.0 eV – 2.1 eV, that is fully acceptable for considering the expansion procedure to be reliable. Since functional group assignment of Peaks 5-7 is rich, a strict analysis of chemical components of the studied ACs responsible for the peaks is rather difficult and requires a cooperative consideration of the issue with analysis of O 1s spectra.

*O 1s spectra*. If a great variety of functional groups provided by MC-OCG is revealed in the case of C 1s spectra at BE over 286 eV, the issue is actively involved in all VFF components of the expanded O 1s spectra. Actually, as seen in Table 7, none of the five VFF peaks can be attributed to simple C=O, C-O and COO groups, which are immersed among other possibilities making the OCG analysis more picturesque. As seen in the table, the peaks FWHMs fill the interval from 0.9 eV to 2.1 eV thus making the performed expansion reliable enough. It should be noted as well that the VFF peak positions after iteration fit quite well the input data.

To analyze the content of OCGs attributed to the studied samples, let us look at the partial contribution of the O 1s VFF peaks into the total intensity of O 1s spectra presented in Fig. 10c. As seen from the figure, the samples differ from each other quite markedly. Thus, the plotting related to ShC is one-humped and reveals a clear dominance of Peak 2. The plotting related to AntX is two-humped and demonstrates the main contribution provided by Peaks 2 and 4. The distribution related to CB624 and CB632 are one-humped and similar each other simultaneously exhibiting the dominant role of Peak 3. Therefore, even this brief analysis gives clear evidence that oxidative products, the formation of which accompanied the consolidation of carbon core structure of the studied amorphics, are different pointing to different chemical history of the products. Following this distribution peculiarities, it is possible to select particular OCGs from the assortment related to each VFF peak that can be attributed to the species.

Actually, as seen in Fig. 10c, the main OCGs related to the ShC are associated with VFF peaks 1 and 2. Analysis performed by Yamada et. al [59] shows that these peculiarities of O 1s spectra are mainly connected with C=O groups, both individual and included in the carboxylic anhydride and lactone compositions (Peak 1) as well as *o*-quinone, aggregated cyclic ether with lactone and carboxyl (Peak 2). In the latter case additional contribution is provided by O=C-**O**-C=O group of anhydride. Moving on to AntX, one may clearly see that Peaks 1 and 2 remain, albeit decreased in intensity, while the main contribution is made by Peak 4. According to the mentioned analysis [59], Peak 4 is attributed to C-**O**-C groups of aggregated cyclic ether with lactone as well as to aggregated cyclic ether and hydroxyl pyran. Obviously, the common species for Peaks 1, 2 and 4 are lactones and aggregated cyclic ethers with lactone that should present the main oxygen contribution into the framing of BSU molecules of AntX. Coming back to ShC, it is evident that lactone as well as aggregated cyclic ether with lactone should be excluded from its OCG content due to very small intensity of peak 4. Reasoning in the same way when moving to CB632 and CB624, it becomes clear that individual C=O groups as well as quinones, lactones and carboxylic anhydrides should be excluded from their OCGs due to practical absence of Peak 1. On the other hand, the dominance of peak 3 favors cyclic ether, aggregated cyclic ether with lactone. Presence of the latter is supported by peaks 2 and 4 while cyclic ether should be considered as the main contributor.

Table 7. Lists of the VFF peaks of C1s and O 1s spectra assigned for oxygen-containing groups and attributed to studied amorphous carbons; expansion data of the samples in terms of PP-FW-RC format (see Fig. 11)

**C 1s spectra[1]**

| VFF peak number | Peak 1 | Peak 2 | Peak 3 | Peak 4 | Peak 5 | Peak 6 | Peak 7 |
|---|---|---|---|---|---|---|---|
| VFF peak top (eV)[2] | 283.6 | 284.3 | 285.0 | 285.4 | 286.5 | 288.3 | 288.7 |
| Possible OCGs on edges[3] | sp$^2$C-H, sp$^2$C influenced by neighboring functional groups | sp$^2$C | sp$^2$C influenced by neighboring functional groups such as C=O, COOR and COOH | C=O | sp$^2$C-OH, O=C-O-C in lactones, pairs of lactones, O=C=C=O of o-quinone, C-O-C of cyclic ethers and aggregated cyclic ethers (ACEs), C-O-C-OH (hydroxy pyran-HP), HP pairs | O=C-OH, O=C-O-C in lactone, pairs of lactones, O=C-O-C=O of acid anhydride, O=C-O-C in ACEs with lactone and in pairs of cyclic ether with lactone | Pairs of lactones, C-O-C-OH in HP, HP pairs, O=C-O-C in ACEs with lactone |
| 1 Graphite GSM2 | - | 0.8eV – 85.01% | - | 1.4eV - 9.1% | 1.3eV - 2.47% | 1.3eV - >0.1% | - |
| 2 Carbon black CB624 | - | 0.8eV – 59.03% | 1.0eV - 9.6% | 1.1eV - 14.37% | 1.2eV - 9.18% | 1.3eV - >0.1% | 1.6eV - 4.25% |
| 3 Carbon black CB632 | - | 0.8eV – 51.54% | 1.1eV - 5.42% | 1.2eV - 24.37% | 1.1eV - 8.9% | 1.7eV - 1.77% | 1.7eV - 2.86% |
| 4 Shungite carbon | 1.1eV - 8.36% | 1.0eV – 67.17% | 1.2eV - 8.51% | 1.3eV - 11.07% | 1.3eV - 4.21% | 1.7eV - >0.1% | 2.1eV - 0.2% |
| 5 Antraxolite | 1.1eV - 8.36% | 1.0eV – 67.27% | 1.2eV - 9.13% | 1.5eV - 10.35% | 1.4eV - 4.4% | 2.0eV - 0.3% | 1.6eV - >0.1% |

**O 1s spectra**

| Peak number | Peak 1 | Peak 2 | Peak 3 | Peak 4 | Peak 5 |
|---|---|---|---|---|---|
| Peak top (eV)[2] | 531.1 | 532.2 | 533.2 | 534.2 | 536.2[4] |
| Possible functional groups on edges[3] | C=O, O=C-O-C=O (acid anhydride), O=C-O-C (lactone), O=C-O-C (pairs of lactones) | O=C-O-C=O (acid anhydride), O=C=C=O (o-quinone), O=C-OH, C= in aggregated cyclic ether (ACEs) with lactone | sp$^2$-OH, C-O-C in cyclic ethers, C-O-C-OH (hydroxy pyran-HP) and pairs of HPs, O=C-O-C(lactone) and pairs of lactones), C-O-C in ACEs with lactone | C-O-C in ACEs, C-O-C-OH in HP and pairs of HPs, C-O-C in ACEs with lactone; C-O-C of pairs of cyclic ethers and a point defect at the basal plane | C-O-C in aggregated cyclic ester |
| 1 Graphite GSM2 | 531.3 eV - 1.3 eV - 0.7% | 532.4 eV - 1.5 eV - 16.4% | 533.4 eV - 1.6 eV - 15.0% | 534.2 eV - 1.7 eV - 51.9% | 535.5 eV – 1.7 eV - 16.0% |
| 2 Carbon black CB624 | 531.3 eV - 1.5 eV - 1.7% | 532.4 eV - 1.6 eV - 25.6% | 533.4 eV - 2.1 eV - 56.5% | 534.4 eV - 1.5eV - 16.0% | 536.4 eV – 1.8 eV - >0.1% |
| 3 Carbon black CB632 | 531.2 eV - 0.9 eV - 1.1% | 532.3 eV - 1.4 eV - 23.9% | 533.3 eV - 1.9 eV - 46.5% | 534.3 eV - 1.7 eV - 28.3% | 536.3 eV – 1.9 eV - 0.2% |
| 4 Shungite carbon | 531.1eV - 1.6 eV - 26.2% | 532.2 eV - 1.9 eV - 43.4% | 533.2 eV - 1.7 eV - 20.8% | 534.2 eV - 1.4 eV - 6.4% | 536.2 eV – 2.1 eV - 3.2% |
| 5 Antraxolite | 531.2 eV - 1.5 eV - 13.5% | 532.2 eV - 1.6 eV - 29.2% | 533.2eV - 1.8 eV - 12.2% | 534.2 eV - 2.4 eV - 34.8% | 536.2 eV – 1.8 eV - 10.3% |

[1] PP-FW-RC format of the studied ACs does not involve VFF peak positions due to fixation of the input peak positions given in the table during expansion.
[2] Reference VFF peak positions are taken from [59].
[3] The peak attribution is taken from [59].
[4] The peak position is added in the current study.

Naturally, the obtained results of the O 1s spectra analysis should be consistent with that one of C 1s spectra. Coming back to C 1s spectra we can see that a large assortment of OCGs associated with each VFF peak allows speaking about general consistency of O 1s and C 1s spectra. The only peculiarity of C 1s spectra is connected with peak 7 at 288.7 eV observed in the spectra of CB632 and CB624. According to the suggested assignment [59], the peak should be associated with O=**C**-O-**C** compositions in aggregated cyclic ether with lactone. The groups do present in the samples, moreover in bigger number in CB632 in comparison with CB624 according to relative intensities of the bodies Peaks 4 in O 1s spectra (see Fig. 10c). As seen in Fig. 11a, this relationship remains in C 1s spectra as well. Therefore, the contribution of the C 1s spectra analysis into the OCG assignment related to the studied ACs is less significant compared to O 1s spectra. Apparently, this feature is characteristic for graphene-like carbon materials. However, the situation changes when oxygenation concerns basal plane of graphene molecules as was shown by analyzing XPE spectra of heat-treated graphene oxide [59]. As for ACs, suggested OCG compositions of the studied samples are listed in Table 8. According to the suggested assignment, shungite carbon can be classified as "C=O" $sp^2$ carbon, antraxolite – as "O=C-O-C" $sp^2$ carbon while "C-O-C" $sp^2$ carbon matches the best carbon black.

**Table 8**. Oxygen-containing-group content related to the studied amorphous carbons
________________________________________________________________

**ShC**:  carbonyls $sp^2$C=O; acid anhydride O=C-O-C=O; *o*-quinone O=$sp^2$C-$sp^2$C=O, carboxyls $sp^2$C=OOH.
**AntX**:  hydroxyls $sp^2$-OH; C-O-C-OH (hydroxy pyran-HP) and pairs of HPs; C=OOC(lactone) and pairs of lactones; aggregated cyclic ether with lactone.
**CB632**:  C-O-C in cyclic ether and aggregated cyclic ether; C-O-C of pairs of cyclic ether and aggregated cyclic ether with lactone.
**CB624**:  C-O-C in cyclic ester, aggregated cyclic ester and aggregated cyclic ether with lactone
________________________________________________________________

Proposed for the first time such a classification of amorphous carbon undoubtedly requires confirmation. Apparently, the latter should be expected from the difference of such global physico-chemical properties of the species as their acidity, water wettability, water retention and so on so that investigations in this field are highly desirable. At the same time, the fact of the different classes of framed oxide graphene molecules was set in experiment on heat-treated graphene oxide [59], which laid the foundation of the suggested approach to the new VFF expansion of XPE spectra. Actually, analyzing the behavior of XPE spectra in the course of heat treating of graphene oxide, the authors found the following regularities. Started at 373K circumference of graphene oxide involves *o*-quinone, acid anhydride as well as cyclic ether and $sp^2$C-OH under question. Heating at 573K provides removing epoxide groups from basal plane possibly leaving behind different defects. As for the circumference, the main place is taken by lactones and cyclic ether with $sp^2$-OH and acid anhydride under question. Lactones still remain the main groups in the area at heating at 773K and are replaced by aggregated cyclic ether with and without lactones at heating at 1223K. The authors particularly pointed a constant ratio of the intensity of C 1s and O 1s spectra at increasing temperature over 573K, that evidences the conservation of the O/C content, while changes in the composition within the framing zone of molecules caused by a reconstruction were observed. Therefore, at a qualitative level, passing from graphene oxide heated at 373k to that one heated 573K-773K is similar to passing from shungite carbon to antraxolite. A further heating at 1273K provides a product similar to carbon black. The revealed similarity may evidence a governing role of the temperature regime of the studied amorphics formation. Actually, both the similarity of the products formed in due course of graphene oxide heat-treating and the studied amorphous carbons and the role of temperature in the latter production need a further scrupulous study.

## 4. BSUs structure and properties

### 4.1. Model structures of BSU of the studied $sp^2$ amorphous carbons

Structure data as well as data on quantitative and compositional chemical content of the studied ACs allow addressing the main goal of the study aiming at the construction of reliable molecular models of the $sp^2$ amorphic carbon BSUs. According to the structure investigation, BSUs of the studied ACs present flat framed graphene molecules of ~2 nm in size (see Table 1). Among a large number of graphene structures that could be proposed as the main model of suitable nanosized graphene molecule, the (5,5)NGr molecule seems to be a good choice. Actually, this molecule, presenting a right-angled graphene sheet with five benzenoid units along armchair and zigzag edges and covering in van-der-Waals atom diameter 1.5x1.5 nm$^2$ of space, has been the main model of numerous quantum-chemical studies of graphene accumulated in monograph [63], which allows considering each its new property on a valuable theoretico-computational background. Twenty two edge atoms accumulate the main portion of the molecule chemical activity and form its circumference in expectance of the molecule chemical modification [64]. These atoms are the main target of hydrogenation and oxidation as well as any other chemical attack occurred in due course of the ACs generation, a quantitative information about which is collected in Table 6. As was previously said, we put aside all other chemical participants of the samples chemical story and will consider carbon-hydrogen-oxygen triad only. The mass content of hydrogen, recalculated in at% and applied to carbon mass of the (5,5)NGr molecule, determines 5 hydrogen atoms for ShC model, while 9, 3 and nil (0.5) in the case of AntX, CB632 and CB624, respectively. The relevant atomic contribution of oxygen, based on XPS study, constitutes 4 atoms for ShC, AntX, and CB632, while 3 atoms for CB624. It is obvious that the numbers present statistical averaging over large number of BSUs thus allowing the deviation to one side or the other from the average value. With this in mind we can suggest the following molecular models of the studied ACs BSUs.

*"C=O" shungite carbon.* Figure 12 accumulates a set of model BSU molecules suggested for consideration. The molecules framing consists of 6 hydrogen and 4 oxygen atoms. The number of hydrogens was made even for the analysis of singlet/triplet multiplicity of the ground state to be possible. Since graphene molecules belong to open-shell ones, the models were equilibrated in the framework of AM1-UHF approximation (see details in [65]). The total energies of the ground state $E_{gr}$ are listed in Table 9 alongside with the energy of the singlet/triplet gap $\Delta E_{ST} = E_{gr}^{sg} - E_{gr}^{tr}$, sign of which points to either singlet (minus) or triplet (plus) ground state, and the total number of effectively unpaired electrons (eue) $N_{gr}$ which present radical efficiency of the molecules. Models in Fig. 12 are grouped in two sets, namely, (a)-(f) and (g)-(l), respectively, once arranged in each set in order of increasing the total energy.

Models (a)-(f) of the first set reflect a possibility to construct framed graphene oxyhydrides with a given number of oxygen and hydrogen atoms following recommendation listed in the first row of Table 8. As seen in Table 9, the choice is actually not so big. Among OCGs attached, the main position, as expected on the basis of detailed study of the graphene molecule oxygenation [65], should be taken by a quartet of carbonyl groups with their least

$E_{gr}$=784.79 kcal/mol. However, this chemically simple composition evidently does not dominate in the O 1s spectrum of the species due to which carboxylic anhydride accompanied by one carbonyl heads the group. Then follow one pair of carbonyls with *o*-quinone and a pair of *o*-quinones that completes the first part of the series of energy friendly derivatives. Addition of a sterically cumbersome carboxylic unit at any stage of oxidation significantly increases $E_{gr}$ making the reaction not favorable even for (e) and (f) models, for which the change is minimal. As was shown [65], carboxyls actually are not friendly addends of graphene molecules due to energetic reason, the fact of which is supported experimentally by a number of reduced graphene studies (see review [66]) pointing to a very small contribution of the latter.

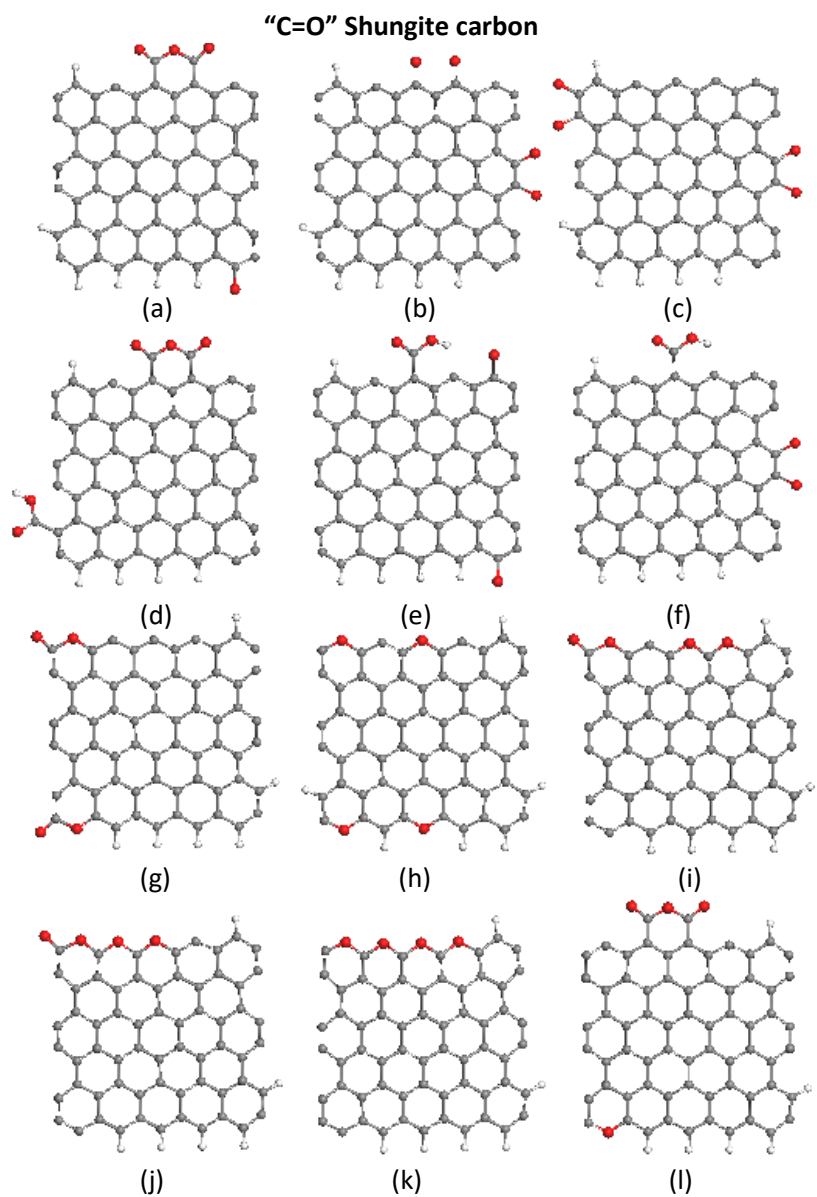

**Figure 12.** Equilibrium structures of molecular BSU models of "C=O" shungite carbon based on (5,5)NGr with a number of OCGs, UHF calculations: (a) carboxylic anhydride and carbonyl; (b) two carbonyls and *o*-quinone; (c) two *o*-quinones; (d) carboxylic anhydride and carboxyl; (e) carboxyl and two carbonyls; (f) o-quinone and carboxyl; (g) two bare lactones; (h) four cyclic ether; (i) a bare lactone and a pair of cyclic ether; (j) aggregated cyclic ether with bare lactone; (k) aggregated cyclic ether; (l) carboxylic anhydride and a cyclic ester.

**Table 9.** Energetic (*kcal/mol*) and radical (*eue*) characteristics of amorphous carbon models[1]

| | Shungite carbon | | | | | |
|---|---|---|---|---|---|---|
| **Models** | *(a)* | *(b)* | *(c)* | *(d)* | *(e)* | *(f)* |
| $E_{gr}$ | 801.27 | 814.07 | 825.28 | 849.20 | 880.91 | 899.09 |
| $\Delta E_{ST}$ | 20.07 | 13.54 | -17.25 | 6.35 | 16.01 | 13.05 |
| $N_{gr}$ | 27.4 | 24.7 | 24.3 | 24.5 | 24.0 | 24.7 |
| **Models** | *(g)* | *(h)* | *(i)* | *(j)* | *(k)* | *(l)* |
| $E_{gr}$ | 769.33 | 777.95 | 782.07 | 790.63 | 803.69 | 828.70 |
| $\Delta E_{ST}$ | -27.61 | 5.91 | 14.23 | -9.94 | -6.91 | 3.90 |
| $N_{gr}$ | 26.1 | 22.5 | 24.6 | 24.4 | 23.6 | 24.6 |
| | Antraxolite | | | | | |
| **Models** | *(a)* | *(b)* | *(c)* | *(d)* | *(e)* | *(f)* |
| $E_{gr}$ | 541.39 | 545.83 | 565.16 | 568.30 | 564.98 | 571.30 |
| $\Delta E_{ST}$ | 19.52 | 12.16 | 35.94 | -7.77 | -10.38 | -7.10 |
| $N_{gr}$ | 22.2 | 22.4 | 21.9 | 21.0 | 19.8 | 20.5 |
| **Models** | *(g)* | *(h)* | *(i)* | | | |
| $E_{gr}$ | 597.41 | 632.16 | 653.23 | | | |
| $\Delta E_{ST}$ | 4.74 | 19.82 | 12.56 | | | |
| $N_{gr}$ | 21.8 | 24.3 | 23.9 | | | |
| | Carbon black CB632 | | | | | |
| **Models** | *(a)* | *(b)* | *(c)* | *(d)* | *(e)* | *(f)* |
| $E_{gr}$ | 919.56 | 945.15 | 955.87 | 956.36 | 957.39 | 962.64 |
| $\Delta E_{ST}$ [2] | - | - | - | - | - | - |
| $N_{gr}$ | 29.7 | 28.1 | 26.5 | 24.2 | 26.0 | 26.5 |
| | Carbon black CB624 | | | | | |
| **Models** | *(a)* | *(b)* | *(c)* | *(d)* | *(e)* | *(f)* |
| $E_{gr}$ | 1112.20 | 1129.02 | 1130.57 | 1141.13 | 1150.84 | 1158.19 |
| $\Delta E_{ST}$ | 11.37 | 9.23 | 11.25 | -13.31 | 6.50 | -8.25 |
| $N_{gr}$ | 30.9 | 30.6 | 30.1 | 27.1 | 27.3 | 27.7 |

[1] See models in Figs. 12, 13 and 14, respectively.
[2] The model ground state is doublet.

The performed analysis does not allow making a clear conclusion which namely models are realized in the case of shungite carbon in practice since the energy $E_{gr}$ is a characteristic of a potential ability only while the nature of empirical observations lies in the chemical kinetics rather than the thermodynamic stability of the products. Reaction occurrence is governed by important kinetic parameters, among which there are chemical reactivity including multiple reaction pathways, isomerization, stereo- and regiospecificity related to reactants and products that are free energy basins separated by barriers of different high. Until now, a possibility to quantitatively consider the relevant problems related to the formation of ShC BSUs, particularly taking into account a large variety of reactants, seemed to be quite impossible. However, the appearance of a new method, called by its authors 'a multi class harmonic linear discriminant analysis' (MC-HLDA) [67], presenting metadynamics with discriminants as a tool for understanding chemistry, inspires great optimism that in the near future similar complex problems can be solved.

Models (g) - (l) in Fig. 12 present a reconstruction of oxygen framing of the ShC BSUs that might be expected as a result of long heating of the body similarly to that observed for graphene oxide [59] and discussed above. Removing of carbonyls, followed by further possible oxygen migration over the molecules circumference, leads to the formation of benzenoid heterocycles of different kinds among which the consistently rising energy $E_{gr}$ corresponds to a pair of lactones, a quartet of cyclic ether, lactone and two cyclic ether, aggregated cyclic ether with and without lactone and so on, respectively. Sufficient number of 'empty' carbon atoms allows constructing bare lactones without any hydrogen atoms. As seen in Table 9, energies $E_{gr}$ of all these derivatives are compared with those related to models (a) – (f), even less concerning the first three ones, causing a question why not (g) – (l) but (a) – (f) models form the ground of ShC BSUs? Apparently, the answer to the question should be looked for in the peculiar kinetics of the reactions involved. Let us hope that the application of the MC-HLDA method to the case will provide a needed detailed understanding of the chemistry of shungite carbon formation.

*"O=C-O-C" antraxolite*. Models (a) – (i) in Fig. 13 present a view of oxygen framing of graphene molecules in the presence of 4 oxygen and 10 hydrogen atoms (even number of the latter is taken to consider singlet/triplet spin multiplicity of the ground state) according to the recommendation listed in the second row of Table 8. In this case, the MC-OCG approach [59] may fit a two-humped distribution of the intensity of VFF peaks in Fig. 11c. As in the case of ShC, the models sequence from (a) to (i) corresponds to continuous increasing of the total energy $E_{gr}$. The main part of this model collection is similar to ShC models (g) – (l) presenting the 'reconstructed' oxygen framing. However, as seen in Table 9, increasing hydrogen atoms number causes a significant decreasing of $E_{gr}$ of AntX models, which reveals a strong effect of hydrogen atoms on the electronic system of graphene molecules well known before [64, 68]. The model succession of 'reconstructed' oxygen framing is changed as well. Thus, the first place in this succession is assigned to a quartet of hydroxyls (not shown in Fig. 13 since it corresponds to only one VFF peak in Fig. 11c, which contradicts to the observation) with $E_{gr}$= 496.82 kcal/mol. The next two places (a) and (b) correspond to pairs of lactones, involving one hydrogen atom (H-lactone) and oxygen atom of the C-O-C groups of which are located either on a pair of zigzag or on zigzag and armchair atoms, respectively. Then follow aggregated cyclic ether without (c) and in the presence (d) of H-lactone and this group of derivatives is completed by two bare lactons (e), both C-O-C groups of which are located on armchair atoms, and H-lactone, accompanied with two cyclic ether (f). The energy change over models (a) – (f) constitutes 30 kcal/mol, but it drastically changes by about the same value for model (g) by inserting one hydroxyl, which causes a replacement of a H-lactone pair in (a) by H-hydroxy pyran and H-lactone composition in (g). The addition of one more hydroxyl to a bare lactone pair in (e), thus transferring it in a pair of bare hydroxy pyrans in (h), causes a doubling of the energy growth. A replacement of two cyclic ether in (f) model by two hydroxyls (see (i) model) still lifts the energy by ~30 kcal/mol. Therefore, models (g) – (i) have less probability to present BSUs of antraxolite than hydroxyl-free models (a) – (f).

*'C-O-C' carbon blacks CB632 and CB624*. Two sets of (a) – (f) models, presented in Fig. 14, in comparison with similar set in Fig. 13, continues to illustrate the influence of the presence of hydrogen atoms in the framing area as well as of their number on the model total energy and energetic sequences of the models. A limited number of edge atoms of the basic molecule essentially narrows a choice of variant structures due to which OCGs sets in Figs 13 and 14 are well similar. In practice, this restriction remains valid for AntX and CB632 due to a comparable size of their BSU molecules. However, the latter is much bigger for CB624, due to which its model presentation in Fig. 14 is not full. This means that due to longer edges the aggregated cyclic ether chains might be longer, their ending by lactone might be less energetically profitable thus explaining a narrowing of CB624 VFF peak in Fig. 11c with respect to that of CB632. As for comparison of AntX and CB632, a drastic difference between their VFF distributions seemingly does not result directly from the sets of models presented in Fig. 13 and 14. However, the subsequences of the energetically favorable models are different in the two cases. The difference inevitably provides significantly different kinetics, which, in its turn, governs the assortment of the final products. In view of this, becomes apparent a very important role of hydrogen atoms, both their number and dispositon over edge carbon atoms. Completed by the above consideration concerning CB624, intuitively grows the confidence that the disclosed importance of hydrogen atoms is not occasional but just vital since seemingly these very atoms provide the termination of any reactions on the edge atoms of growing graphene sheet, thus determining size of the latter [6, 69]. Accordingly, it might be possible that very low hydrogen content of CB624, which is close to statistical zero, but not exact nil, determines the size of its BSUs that is much bigger than that of the other amorphics.

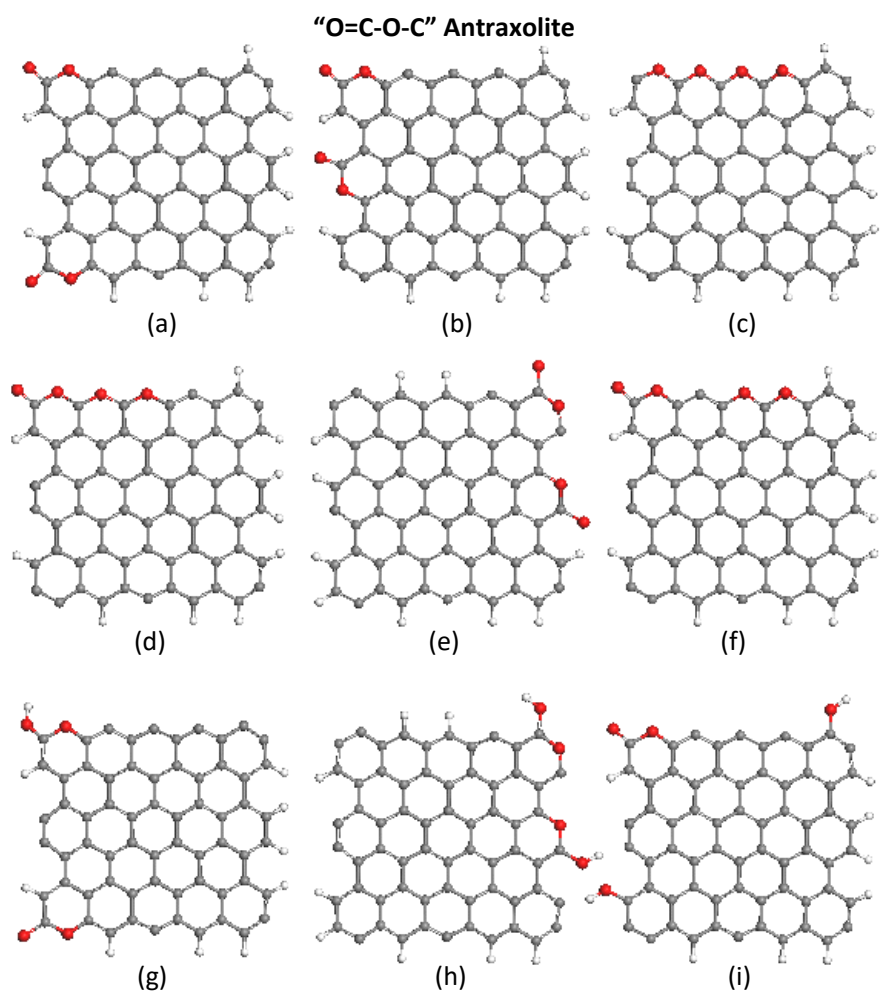

**Figure 13.** Equilibrium structures of molecular BSU models of "O=C-O-C" antraxolite based on (5,5)NGr with a number of OCGs, UHF calculations: (a) two H-lactones at zigzag edges; (b) H-lactone and a bare lactone at zigzag and armchair edges, respectively; (c) aggregated cyclic ether; (d) aggregated cyclic ether with H-lactone; (e) two bare lactones at armchair edges; (f) H-lactone and a pair of cyclic ether; (g) two H-hydroxy pyrans at zigzag edges; (h) two bare hydroxy pyrans at armchair edges; (i) H-lactone and two hydroxyls.

"C-O-C" Carbon black
CB632

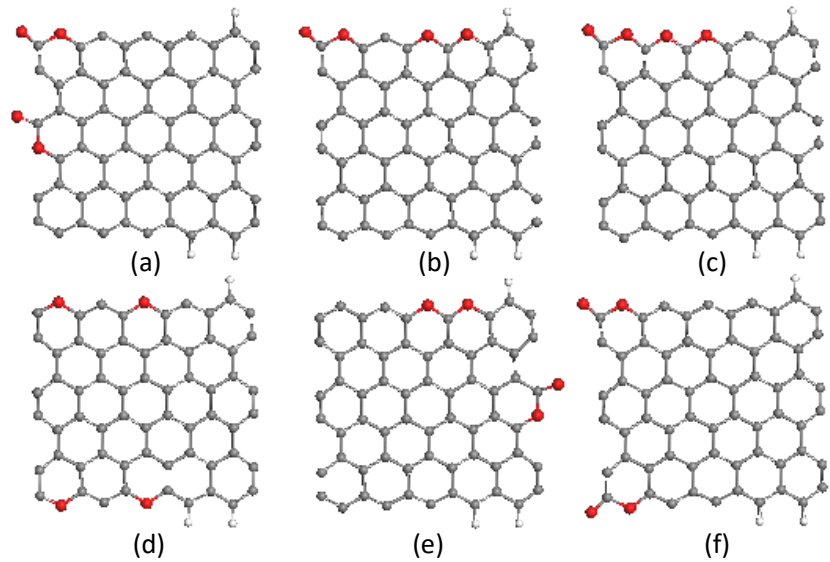

CB624

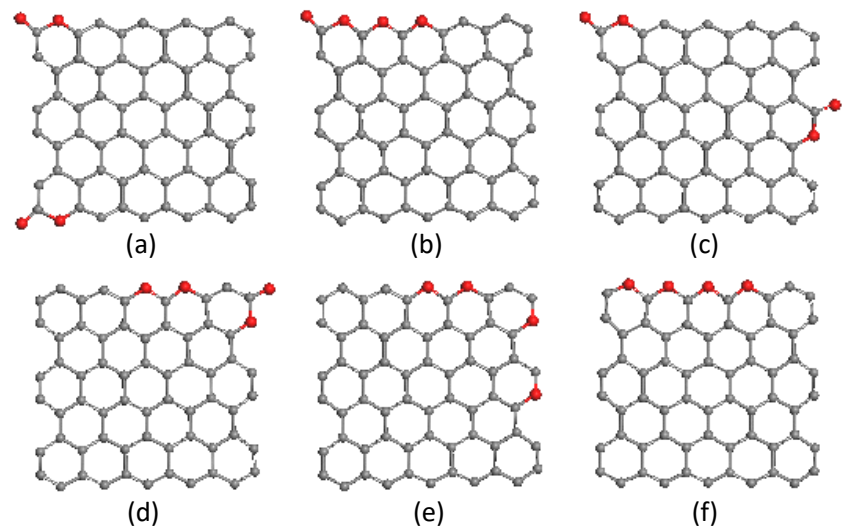

**Figure 14.** Equilibrium structures of molecular BSU models of "C-O-C" carbon blacks based on (5,5)NGr with a number of OCGs, UHF calculations. CB632: (a) two bare lactones at zigzag and armchair edges; (b) a bare lactone at zigzag edge and a pair of cyclic ether; (c) aggregated cyclic ether with a bare lactone; (d) four cyclic ether; (e) a bare lactone at armchair edge and a pair of cyclic ether; (f) two bare lactones at zigzag edges. CB624: (a) two bare lactones at zigzag edges; (b) aggregated cyclic ether with a bare lactone; (c) two bare lactones at zigzag and armchair edges; (d) a bare lactone at armchair edge and a pair of cyclic ether; (e) a pair of cyclic ether and two cyclic ether at zigzag and armchair edges, respectively; (f) aggregated cyclic ether.

### 4.2. Common characteristics of model BSUs

The collection of models presented in Figs. 12-14, shows quite detailed what can happen in the circumference of a framed oxyhydride graphene molecule under permissible migration of oxygen atoms while maintaining their total number. And although such behavior was directly observed empirically only in the case of rGO, produced in the course of heat-treating of graphene oxide [59], there is reason to expect such a behavior of any graphene molecules under similar conditions. Evidently, the composition of final products is determined by the chemical surrounding and temperature. From this viewpoint there is a high temptation to firstly connect the difference in the atomic properties of BSUs of the studied ACs with the temperature conditions of their generation. In the case of anthraxolite and carbon blacks such an approach seems to work quite well. Actually, AntX generation is associated with temperature regime 200-400$^0$C [11]. In contrast, carbon blacks are products of high temperature pyrolysis (T~ 700$^0$C – 1200$^0$C) [39]. In both cases the "heat-treated-graphene-oxide" concept is able to simulate the difference in their compositions quite well. As for shungite carbon, the explanation of its BSUs composition peculiarities in the framework of the above concept confronted with the inconsistency of the temperature regime of its generation: the temperature would have to be much lower than that of antraxolite while in practice the two temperatures are practically the same [6, 11]. Apparently, the inconsistency is caused by the difference in chemical surrounding related to the generation of both mineral substances. Anyway, results of the current study seem to be able to stimulate the intensification of the search for the clarification of the 'temperature-surrounding' issue concerning not only these two products, but other natural $sp^2$ carbons.

The second situation, which deserves special attention, concerns the radical properties of the suggested models. As well known [63, 64], graphene molecules, both bare and framed in due course of the first-stage chemical modification, are characterized by high level of radicalization. The latter is adequately described by the total number of effectively unpaired electrons $N_{gr}$. The presented models are not exclusion to the general rule, which is evidenced by $N_{gr}$ listed in Table 9. Actually, the averaged values constitute 25 e, 22 e, 26 e, and 28 e for ShC, AntX, CB632, and CB624 models, respectively. Of these units, in all the cases, 12-13 refer to atoms in the basal plane while the remaining determines chemical activity of the molecules provided by edge atoms. Figure 15 presents equilibrium structure of four models, which head the relevant series in Figs. 12-14 related to the studied ACs together with their chemical portraits written by the $N_{gr}$ distribution over atoms (so called $N_{DA}$-maps [63, 64]). As seen in the figure, black balls of the portraits in the circumference areas of the molecules clearly exhibit a fully inhibited chemical ability of both heteroatoms and carbon atoms to which the former are attached. Nevertheless, the remaining part of the molecules circumferences still remains highly active, particularly in the case of CB624, and easily accessible to not only gaseous reagents, but to bulky ones as well, in spite of aggravation with sterical constrains in the latter case. Therefore the presented models are radicals.

Due to high radicalization of each suggested model and since the models, in general, are fully consistent with the discussed structural-compositional characteristics of the studied ACs, we have to make an astonishing conclusion that $sp^2$ amorphous carbons are conglomerates of stable radicals. The conclusion becomes still more impressive on the background of existing presentations that stable radicals, which are usually presented by quite small molecules, are not ordinary part of the modern chemistry but are rare exclusion, the nature of which has still remained unclear (see reviews [70-73] and references therein). On the other hand, it would seem that the huge amount of $sp^2$ amorphous carbons, both natural and engineered, as well as their exclusively broad involvement in various chemical and technical processes for about a thousand of years could not disregard such an important characteristic of the substance if it existed. Nevertheless, in fact, by now there have not been communications joining radical terminology and huge massive of the available data concerning $sp^2$ ACs. At the same time, specific peculiarities, concerning heterogeneous catalysis and electron-spin resonance, have been really observed and widely used. However, until now they have not been connected with specific radical properties of $sp^2$ carbons as well.

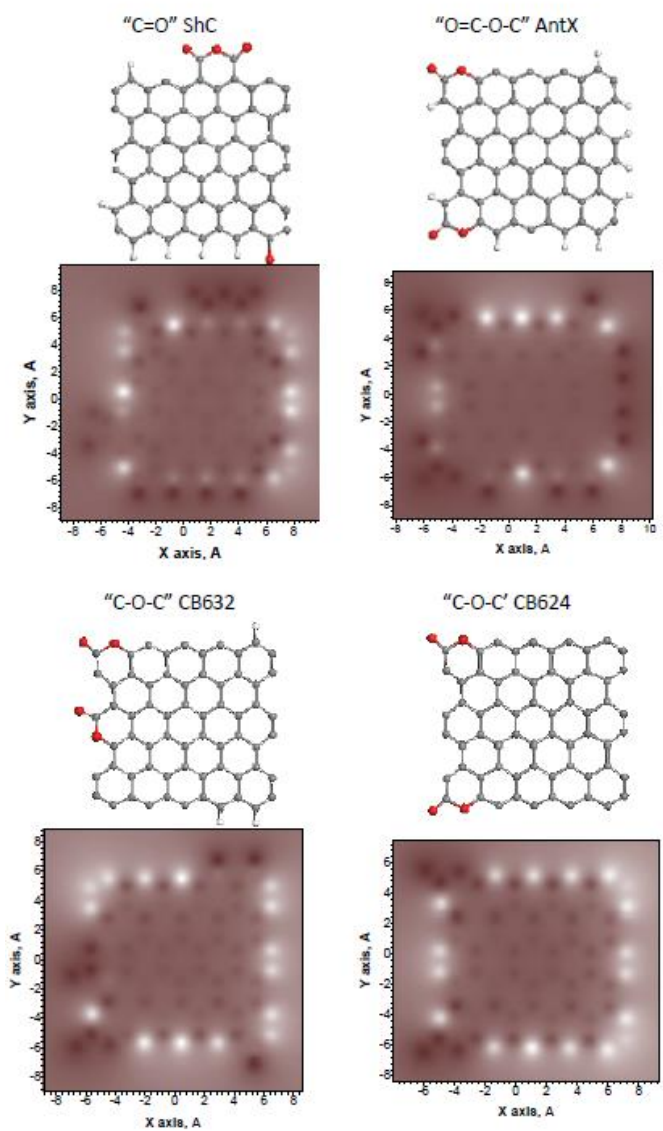

**Figure 15**. Equilibrium structure and $N_{gr}$ distribution over atoms of BSU model molecules (see text).

According to L. Radovich, one of the most reputable expert in the field, "catalytic applications of carbon materials are as old as the discipline of physical chemistry, and probably even older" [74]. Carbon materials such as activated carbons, carbon blacks, graphite, and graphitic materials have been used for ages in heterogeneous catalysis, as either catalysts or catalyst supports. In a large number of publications, thoroughly reviewed in a set of monographs [60-62, 75, 76], carbon materials relevant to catalysis are discussed with a special emphasis given to the description of carbon surface features. A review [77] can be recommended to enter the field where particular attention was given to surface functional groups, mainly OCGs, which are considered to be responsible for the catalytic activity of the carbons. As for natural $sp^2$ carbons, without touching the coals, information concerning their reactivity and catalytic ability is much scarcer, however, despite this, quite convincing. This concerns mainly shungite carbons [78-81], for which a decisive role of their BSUs for these physicochemical processes was clearly shown. Similar investigations of anthraxolites become a top point for the near future.

As far as understanding that $sp^2$ carbon materials possess catalytic properties, ideas about the decisive role of surface functional groups in these processes were transferred to fullerenes, carbon nanotubes, and graphene [62, 82, 83]. However, as seen in Fig. 15, these groups completely lose their activity after attaching to the carbon core and are not capable to further lead any chemical reaction while neighbor 'empty' carbon atoms are ready to play the role. The ignoring of the fact (in contrast to the case of 'legitimate' small stable radicals [71]) should be apparently connected with the way of progress in understanding that chemical composition of BSU of $sp^2$ carbons is responsible for their catalytic activity. Obviously, the BSUs atomization has become possible in the framework of computational QCh. At this point the considerations of small stable radicals and $sp^2$ carbons proceeded in different directions. The former, once small by size, were considered by using high level QCh approaches, mainly based on Hartree-Fock approximation, that take radical properties into account [71]. The latter, once large in size, were considered by using much simpler approaches, predominantly various versions of DFT. However, this approach is spin inactive and losses radicals substituting them by ordinary closed-shell molecules (see a profound discussion of the issue in [63, 84]). At the same time, the dictate of DFT-based virtual QCh, particularly strengthened by the current graphene science, is very strong, which does not allow raising the curtain and bringing the main actor of catalysis - stable radicals to the scene. Against this background, a study [85] that voiced the term "radical" in relation to $sp^2$ amorphics for the first time when attempting to explain carbonaceous soot inception and growth in term of resonance-stabilized hydrocarbonradical chain reactions, is quite revolutionary.

Comparing with a huge number of studies on heterogeneous catalysis involving $sp^2$ carbons, ESR studies seem almost inconspicuous. Nevertheless, the conducted investigations allowed establishing two commonalities characteristic of $sp^2$ carbon: 1. ESR signals as evidence of radical nature of the species were observed in all the cases related to $sp^2$ nanocarbons, namely: graphite, all graphene-based materials, including various soots, coals and other $sp^2$ amorphous carbons, as well as fullerenes and carbon nanotubes, shungite carbon and antraxolite (see [80, 86-90] but a few). 2. ESR signals in all the studied cases are well similar being presented by a broad single peak and evidencing the same specific peculiarity of the spin system of the studied samples. Evidently, the spin density delocalization over all $sp^2$ carbon atoms of the species (see an example for $sp^2$ amorphics in Fig. 15 as well as for various graphene systems [63], fullerenes and carbon nanotubes [91]) is responsible for the specificity. Apparently this feature may explain a drastic inconsistency concerning the intensity of EPR signals of amorphous carbons. The estimations, based on a standard concept of localized spins in both the studied samples and referent standards, give values of $10^{16}$-$10^{17}$ spins/g in all the cases. Since 1 gram of the carbon body composed of, say, (5,5)NGr molecules, each of which involves 66 carbon atoms, consists of ~$10^{21}$ atoms, the measured spin concentration implies the presence of unpaired electrons, whose number differ from that of atoms by 4-5 orders of magnitude. At the same time, as seen from Table 9, the ratio for the suggested model constitutes ~1/3 that is critically inconsistent with the measured values. Seemingly, this inconsistency has a direct relationship with peculiarities of magnetic properties of open-shell electronic systems [92-94] concerning spin alignment, Hund's multiplicity rule, dynamic spin polarization, and so forth, all of which fully changes the approach to the spin concentration determination.

Restoration of the BSUs radicals in their rights confronts us with new problems to be solved and the first one is to understand why the *sp²* carbon molecular radicals are stable or, by other words, which prevents the completion of residual valence inhibition of *sp²* carbon atoms in the circumference and defect areas (see bright balls in Fig. 15). The issue has been a top problem since the first steps in graphene chemistry (see review [95] and references therein). Three main factors have been disclosed since that favor the stabilization of radical character of open-shell graphene molecules: 1. Spin-delocalized character of the molecule radicalization provided by the conjugation of *sp²* electrons over the total number of carbon atoms; 2. Triplet spin multiplicity of the molecules ground state; 3. The presences of heteroatoms located on edge atoms of the molecules. All these three factors are characteristic for the suggested BSUs models, as seen in Table 9 and Fig. 15. However, they are only consequences of the spin topochemistry of spatially extended *sp²* nanocarbons in general and graphene molecules in particular [63, 96, 97], in the frame work of which the answer to raised question should be searched for.

Topochemistry of *sp²* nanocarbons is quite specific and reveals not only dependence of the final products composition on size and shape of the carbons in use, but on their chemical prehistory, particularly concerning edge and ending atoms of graphene and nanotubes. A lot of information on the issue is collected in monographs [60-62, 76] and reviews [77, 82, 83] related to heterogeneous catalysis of metal-free *sp²* carbons. Attributing to heteroatoms, it was shown that each of them writes its own story. The feature is typical to graphene topochemistry [63] due to which not only oxygen and hydrogen, but nitrogen, sulfur and other heteroatoms actively participate in the formation of massive *sp²* ACs. As seen in Table 6, these atoms contribution is comparable with that of related to oxygen-hydrogen couple. Consequently, the oxyhydride picture of the studied amorphics BSUs is not complete and its extending over tio- and nitro-composition becomes the main goal of the further study.

5. **Conclusive remarks**

Concluding the presentation of obtained results and their analysis, it remains to show what normal (old), expected and new, announced in the paper title, can be suggested for the further consideration. Let us consider all these three issues related to the paper fractions given above.

*Sample characterization*. It occurs normal, that all the studied ACs are multielemental bodies with a predominant contribution of carbon. Expected is the finding that ACs are supplemented with hydrogen, oxygen, sulfur, nitrogen and other minor impurities. What is new that none of the studied elitist ACs, including C-standard, destined for a quantitative C-element analysis by EDS, as well as specially selected graphites, are carbon pure. All of them include 1-3 wt% of oxygen. Natural ACs contain as well up to a few wt percents of hydrogen. Accordingly, such mapping as "C 99.95 wt%", which can be often met at websites of respected carbon producers, is an obvious overestimation.

*Analytical techniques ability*. It turns out to be normal that none of the considered analytical techniques provides a correct determination of the chemical content since actually all of them 'see' different objects. Expected is that the quantitative estimation of weight contribution of a selected element into the sample mass is highly uncertain. New is that addressing, say, C-content, TGA offers a 'brutto' estimation under the term of 'fixed carbon' that covers all other chemical elements that are not released as ash, among which there are so important additives as hydrogen, oxygen, sulfur. Widely used HCNS EA significantly overstates the carbon content not taking into account oxygen and other impurities besides title-declared while so called minor impurities may constitute a few percents, particularly for natural species.

Expected is that, oppositely to HCNS EA, EDS and XPS do overestimate the carbon content due to not taking into account hydrogen whose contribution may be considerable. Relatively to ACs, both techniques are mainly considered as "O-tools". New is that thus determined oxygen content differs a few times in favor of XPS. The latter is explained by particular spatial anisotropy of graphene molecules in view of which a small escape depth of the XPE flux acts as a spatial selector of particular molecule zone but not a whole molecule.

*Molecular structure of amorphous carbons.* Normal is that according to microscopic study ACs are nanostructured objects which allows considering their structure in terms of nanosize basic structural units. Expected is that the BSUs present framed graphene molecules of nanometer in size. New is that the BSUs sizes form the series AntX < ShC ≤ CB632 << CB624. The series is suggested to be explained by the inverse power series of hydrogen content in the samples, putting the latter to be responsible for termination the growth of graphene sheet in time.

Normal is that, proven by structural analysis, BSUs of the studied ACs represent framed graphene molecules with expectedly complicated chemical composition of their circumferences. Actually, quite long ago has been found empirically that additives of carbon black are located at the periphery of benzoid-arranged planar sheets [75]. Quantum-chemical explanation of the issue, based on numerous experimental facts, has been suggested afterwards [64]. New is the suggestion of BSUs models, nanometer in size, basing on the two main concepts as well on the empirical study of ACs chemical content in the current study. Once limited to contributions of hydrogen and oxygen among other minor impurities, the models were composed in the framework of the C-O-H triads. Hydrogen atoms were involved in the BSU framing structure via a set of C-H bonds, the number of which corresponds to the total hydrogen atomic content obtained in the current study. The composition of oxygen-containing groups (OCGs) was established by detailed analysis of the sample XPE spectra in the framework of multi-component OCG approach [59]. The approach allowed dividing the studied ACs into three groups, attributing ShC to "C=O" amorphics, AntX – to "C=O-C-O" ones and both carbon blacks to "C-O-C" species. A limited number of possible model BSU structures were suggested for each kind of the amorphics. A qualitative conformity of the suggested structures with empirically observed for reduced graphene oxide, which was produced under heat-treating of graphene oxide [59], was established and a promising way to connect the peculiarities of molecular structure of the studied ACs with details of the temperature regime of the sample production was proposed. The paper pays special attention to a new observation concerning radical essence of BSUs of the studied ACs, which makes the latter a huge-mass source of stable radicals and which greatly changes the available representation on these solids.

When the article was already finished, the publication [98] of T. Bandosz became known to one of its authors (E. Sh.). One of the authoritative experts on engineered $sp^2$ amorphous carbons, Prof. Bandosz suggested a particular view on old, more than a thousand years in practice, and new, started at the beginning of the nanocarbons (fullerenes, carbon nanotubes and graphene) era, histories of $sp^2$ carbons and their interrelation. This view proved to be deeply in tune with the basic idea of the present study: to look at seemingly well known objects such as char, carbon blacks, activated carbon and so forth from the position of new nanotechnological graphene science for a fresh, out-of-the-box thinking to be brought. The first attempt seems to be quite successful. Of course, the obtained results still leave a lot of questions to be answered and a lot of approaches to be improved. Nevertheless, a fresh air of a new vision of old and highly reputed problems is quite inspiring.

**Acknowledges**

The authors would like to thank I. A. Domashnev and Ye. M. Tropnikov for assistance in performing HCNS EA and EDS measurements, M. F. Budyka and A. V. Naumkin for fruitful and stimulating discussions. The study was carried out within the research topics of Institute of Geology of Komi SC of RAS (GR No. AAAA-A17-117121270036-7) (Ye.G.), project N 212 of the Institute of Geology of Karelian Research Centre of RAS using the equipment of the Core Facility of the IG KarRC RAS (N.R.) , the «RUDN University Program 5-100» (N.P., V.P., E.Sh.).

**Conflict of Interests**

The authors declare that there is no conflict of interests regarding the publication of this paper.